\documentclass[mnsc,nonblindrev,copyedit]{informs3a}
\usepackage{preamble}

\begin{document}
\TITLE{\Large Supracompetitive Pricing Under AI Monoculture}

\ARTICLEAUTHORS{
  \AUTHOR{Shengyu Cao}
 \AFF{Rotman School of Management, University of Toronto, \EMAIL{shengyu.cao@rotman.utoronto.ca}}
 \AUTHOR{Ming Hu}
 \AFF{Rotman School of Management, University of Toronto, \EMAIL{ming.hu@rotman.utoronto.ca}}
}

\ABSTRACT{
When competing sellers delegate pricing to a shared AI model, such as a large language model, the shared model can create correlated recommendations across firms. Combined with performance-driven updates that aggregate seller feedback, this raises the question: Can standard AI deployment practices inadvertently produce supracompetitive pricing?
We develop a stylized duopoly model in which two sellers receive pricing recommendations from a shared AI model characterized by two parameters: a propensity parameter that captures the model's tendency to set high prices, and an output-fidelity parameter that measures the alignment between this tendency and the model's outputs. The model updates its propensity through periodic retraining on observed outcomes. We find that the seemingly prudent practice of configuring AI models for robustness and reproducibility can lead to supracompetitive pricing via a phase transition. We establish a critical output-fidelity threshold: below it, competitive pricing is the unique stable outcome. Above it, the model exhibits bistability, with both competitive and supracompetitive pricing being locally stable and the realized outcome determined by the model's initial propensity. Supracompetitive pricing leads to higher average prices, but occasional low-price recommendations complicate detection. When output fidelity is perfect, full price coordination emerges from any interior initial propensity of the model. For finite training batches of size $b$, when the initial propensity lies in the supracompetitive basin, the probability of supracompetitive pricing approaches 1 as $b$ increases. The region of indeterminate outcomes shrinks at a rate $O(1/\sqrt{b})$, so that larger batches convert more initial conditions into predictably supracompetitive trajectories.
Our results identify a critical fidelity threshold below which competitive pricing is the unique long-run outcome. Any factor that reduces the effective alignment between the shared model's propensity and sellers' actual pricing, whether through diversifying AI providers, introducing recommendation noise, or reducing seller adherence, pushes the market toward competitive outcomes. Larger training batches amplify the risk of supracompetitive pricing by suppressing stochastic fluctuations that might otherwise restore competitive pricing.
}
\maketitle

\section{Introduction}

The rapid adoption of algorithms in commercial decision-making has attracted increasing regulatory scrutiny. In March 2024, the Federal Trade Commission (FTC) and the Department of Justice (DOJ) issued a joint statement warning that algorithmic pricing systems can facilitate illegal collusion, even in the absence of explicit agreements among competitors.\footnote{FTC Press Release, ``FTC and DOJ File Statement of Interest in Hotel Room Algorithmic Price-Fixing Case,'' March 28, 2024.} Subsequent enforcement signals have only intensified: in August 2025, DOJ Assistant Attorney General Gail Slater announced that algorithmic pricing investigations would significantly increase as deployment of the technology becomes more prevalent.\footnote{MLex Market Insight, ``Algorithmic pricing probes to increase as use grows, US DOJ's Slater says,'' August 11, 2025.}

This regulatory concern is particularly salient as sellers increasingly rely on large language models (LLMs) for pricing decisions, an area the regulation has not yet addressed. A 2024 McKinsey survey of Fortune 500 retail executives found that 90\% have begun experimenting with generative AI solutions, with pricing and promotion optimization emerging as a priority use case.\footnote{McKinsey \& Company, ``LLM to ROI: How to Scale Gen AI in Retail,'' April 2024.} In Europe, 55\% of retailers plan to pilot generative AI-based dynamic pricing by 2025, building on the 61\% who have already adopted some form of algorithmic pricing.\footnote{Valcon, ``Dynamic Pricing Predictions for 2025,'' January 2025.} Industry applications are already operational: \citet{wang2025llp} report that Alibaba has deployed an LLM-based pricing system on its Xianyu platform to provide real-time price recommendations for sellers.

This heightened regulatory pressure coincides with remarkable market concentration in the LLM industry. As of 2025, ChatGPT commands approximately 62.5\% of the business-to-consumer AI subscription market, while 92\% of Fortune 500 companies report using OpenAI products.\footnote{Data from Earnest Analytics via Backlinko, ``ChatGPT Statistics 2025.'' The Fortune 500 statistic was announced by OpenAI in April 2024.} This confluence of concentrated AI infrastructure and heightened antitrust concern raises a fundamental question: When competing sellers delegate pricing to a shared AI model, can the model's learned pricing propensity produce supracompetitive pricing outcomes, even without any communication between sellers?

The possibility that algorithms might learn to collude has been studied extensively in the context of machine learning/reinforcement learning (RL). \citet{banchio2022artificial} demonstrate that Q-learning agents can converge to supra-competitive prices in repeated games through trial-and-error learning. However, such collusion typically requires extensive training periods and millions of interactions, providing some reassurance that computational barriers might limit practical concern. LLMs present a fundamentally different paradigm. Unlike RL agents that learn pricing strategies from scratch through repeated market interactions, LLMs arrive pretrained on vast corpora of human knowledge, including business strategy, economic theory, and pricing practices. Recent experimental work by \citet{fish2024algorithmic} shows that LLM-based pricing agents converge to supra-competitive prices much more rapidly than their RL counterparts: GPT-4 agents reach near-optimal collusive pricing within 100 periods, compared with the hundreds of thousands to millions of periods required by Q-learning algorithms. Yet the structural mechanisms, beyond in-context strategic reasoning, through which a shared AI model might sustain elevated prices remain underexplored.

\begin{figure}[!htb]
    \centering
    \includegraphics[width=.9\linewidth]{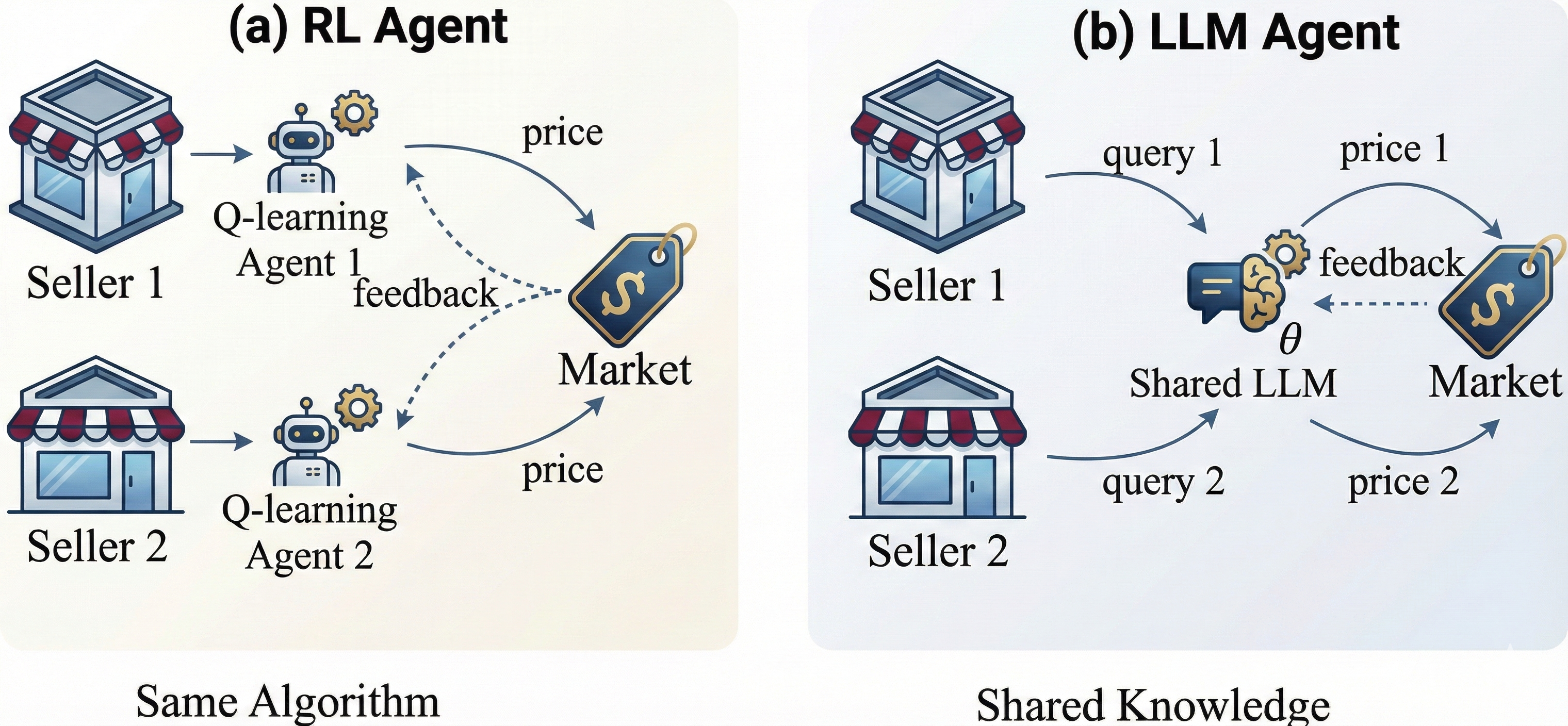}
    \caption{Comparison of pricing mechanisms. (a) Under reinforcement learning, each seller deploys an independent algorithm that learns pricing strategies from scratch through repeated market interactions. Collusion emerges slowly over millions of iterations. (b) Under shared-AI pricing, both sellers query a common pretrained model with a latent high-price preference $\theta$. The shared knowledge infrastructure generates correlated recommendations, while data sharing during retraining aggregates seller feedback, creating a self-reinforcing loop that can drive convergence toward supracompetitive outcomes.}
    \label{fig:rl-vs-llm}
\end{figure}

We identify two sources of supracompetitive pricing risk in LLM-based decision-making, and illustrate the contrast with independent RL-based approaches in Figure~\ref{fig:rl-vs-llm}. Specifically, the first source is the \emph{shared knowledge} infrastructure that arises from market concentration. Given the dominant position documented above, competing sellers frequently delegate pricing decisions to the same AI model. When multiple sellers query the same model, their adopted recommendations may become correlated through the model's internal latent preferences, a structural tendency toward certain pricing strategies encoded in its weights. This correlation creates implicit coordination without any communication between sellers: even if each seller queries the LLM independently, the common underlying preference induces positive correlation in their pricing actions. Moreover, shared knowledge persists even when sellers use different LLM providers. Leading LLMs are often distilled into smaller models by competitors, thereby transferring embedded pricing heuristics from dominant providers to the broader ecosystem.\footnote{For instance, many open-source models, such as LLaMA derivatives, are trained to mimic the outputs of frontier models like GPT-4.} In addition, pretraining corpora across providers draw from largely overlapping public sources, including business literature, economic textbooks, and strategic discussions that extensively document the profitability of coordinated pricing. Alignment procedures such as reinforcement learning from human feedback (RLHF) further homogenize model behavior, as evaluators or users at different companies apply similar standards of ``helpfulness'' and ``reasonableness.'' This raises a fundamental question: Does shared knowledge infrastructure lead to correlated pricing that produces supracompetitive outcomes?

The second source is \emph{performance-driven model updating}. AI pricing models may be periodically retrained or fine-tuned based on user outcome signals. While providers do not directly observe seller profits, they increasingly collect interaction data that carries outcome-relevant information: conversation logs, user satisfaction feedback (e.g., thumbs-up/down ratings), continued adoption versus abandonment of recommendations, and explicit data-sharing arrangements. Major providers retain such data by default.\footnote{Google's Gemini uses a ``Keep Activity'' setting enabled by default, allowing the platform to analyze conversations for AI training. Anthropic's August 2025 policy change similarly made data sharing the default setting.} Enterprise deployments routinely fine-tune models on business-specific performance signals, and platform providers such as Alibaba have deployed end-to-end AI pricing models that directly observe seller outcomes \citep{wang2025llp}. Even coarse outcome signals, such as whether a seller continues to follow pricing recommendations or starts overriding them, reveal whether recommended prices are working. When such signals aggregate across competing sellers during retraining, they create a self-reinforcing loop: pricing patterns associated with continued user engagement are reinforced in future recommendations. This raises a fundamental question: Does performance-driven updating of a shared AI model create a feedback loop that drives prices toward supracompetitive levels?

This paper develops a stylized theoretical model to analyze when a shared AI model's pricing propensity can yield supracompetitive outcomes, thereby addressing the two questions raised above. We find that supracompetitive pricing can emerge as an unintended consequence of standard AI deployment practices. In high-stakes tasks such as pricing, decision-makers typically configure LLMs for robustness and reproducibility by setting decoding temperatures to zero or near-zero. This seemingly prudent choice increases the correlation between the model's internal preference and its generated outputs, thereby facilitating coordination without explicit communication. Moreover, the scale of AI deployment means that each retraining cycle aggregates interactions from many sellers across many pricing rounds, producing large effective batch sizes that suppress beneficial noise, which might otherwise allow prices to move to competitive levels.

{\bf Model.} We consider a symmetric duopoly pricing game in which two sellers compete by simultaneously choosing between a high-price strategy ($H$) and a low-price strategy ($L$). Both sellers delegate their pricing decisions to a shared AI model (motivated by the concentration of LLM providers documented above). The model's behavior is characterized by two parameters: a \emph{propensity parameter} $\theta \in [0,1]$ representing the model's internal preference for high-price recommendations, and an \emph{output fidelity} $\rho \in [0.5,1]$ capturing the alignment probability between this preference and the generated output. The model updates its propensity through model retraining: after observing actions and outcomes over every batch of $b \geq 1$ decision rounds, the model evaluates the relative performance of high- versus low-price recommendations and adjusts $\theta$ accordingly. This update follows a log-odds recursion that increases $\theta$ when high prices outperform and decreases it otherwise.

{\bf Results.} We first analyze a benchmark setting in which model retraining is performed on large batches of interaction data, so that estimation noise becomes negligible. We show that whether high-price recommendations become self-reinforcing depends on two competing forces: the benefit of coordination when both sellers receive and follow the same high-price recommendation, versus the cost of miscoordination when sellers receive different recommendations and the low-price seller captures the market. When output fidelity is sufficiently high, coordination benefits dominate, and learning reinforces the model's propensity toward high prices.

It is worth noting that high output fidelity is precisely what decision-makers seek in practice. Low fidelity, corresponding to high-temperature decoding, introduces randomness into recommendations. Such unpredictability imposes operational costs on sellers: inconsistent pricing confuses customers and erodes brand equity, noisy outputs obscure whether realized profits stem from sound strategy or mere chance, and stochastic recommendations undermine the explainability that regulators increasingly demand of algorithmic pricing systems. Consequently, sellers naturally configure AI models for robustness and reproducibility by favoring low-randomness settings, which pushes output fidelity to high levels.

We establish a phase transition in long-run pricing behavior, governed by a critical output-fidelity threshold. Below this threshold, competitive pricing is the unique stable outcome: the model converges to recommending low prices regardless of its initial propensity. Above this threshold, two stable outcomes coexist: competitive pricing and supracompetitive pricing. Which outcome is realized depends on the model's initial propensity. When output fidelity is perfect, meaning the model's recommendations perfectly reflect its pricing propensity, the model converges to full price coordination from any interior starting point. The standard practice of configuring AI models for reliability thus inadvertently pushes sellers into the parameter region where supracompetitive pricing can emerge.

We then analyze a more realistic setting in which the model is retrained on finite batches, so that randomness persists throughout learning. We show that the system still converges to a stable outcome, but the specific outcome depends on the random training data encountered along the way. Identical models with identical initial conditions can reach different long-run outcomes. We characterize how the probability of supracompetitive pricing depends on batch size and initial conditions. When the model's initial propensity lies in the region that leads to supracompetitive pricing, larger batch sizes increase the probability of supracompetitive pricing by suppressing random fluctuations that might otherwise push the model toward competitive pricing. The zone of genuine uncertainty, where either outcome is possible, shrinks as batch size increases. This result implies that the large-scale deployment of shared AI models, in which each retraining cycle aggregates interactions from many sellers, inadvertently amplifies the risk of supracompetitive pricing due to large effective batch sizes.

\section{Literature Review}
Our work is closely related to three streams of literature: algorithmic monoculture and shared infrastructure, algorithmic collusion, and LLM-based pricing and strategic behavior.

{\bf Algorithmic monoculture and shared infrastructure.}
\citet{kleinberg2021algorithmic} formalize the risks of algorithmic monoculture, showing that widespread adoption of identical algorithms can reduce decision quality by causing all sellers to miss the same opportunities. \citet{bommasani2022picking} demonstrate that sharing model components leads to outcome homogenization: individuals rejected by one decision-maker face rejection everywhere, creating systemic exclusion. \citet{black2022model} show that model multiplicity, the existence of equally accurate models with different predictions, naturally provides diversity across decision-makers, but monoculture eliminates this protective heterogeneity. \citet{kim2025correlated} provide large-scale empirical evidence that even LLMs from different providers and architectures exhibit highly correlated errors, suggesting that provider diversity alone does not eliminate monoculture risks.
Our work demonstrates that monoculture in AI-based pricing creates a direct channel for supracompetitive outcomes: when sellers share the same model, they inherit the same pricing propensity, and performance-driven updating can reinforce that propensity over time.
Our mechanism differs from the omitted-variable-bias channel identified by \citet{hansen2021algorithmic}, in which independent misspecified bandit algorithms produce correlated pricing experiments that drive up prices. In their setting, each seller runs a separate algorithm, and correlation arises from common misspecification. In our setting, correlation is structural: a single shared model issues recommendations to all sellers from a common pricing propensity, and performance feedback from multiple sellers reinforces this propensity over time.

{\bf Algorithmic collusion.}
A growing body of research examines whether pricing algorithms can sustain supra-competitive prices without explicit coordination (see \citealt{Bichler2025AlgorithmicPricing} for a survey). \citet{hansen2021algorithmic} show that when competing sellers independently run misspecified bandit algorithms, correlated pricing experiments create an omitted variable bias that drives long-run prices to supra-competitive levels, even when algorithms do not observe competitors' prices. In reinforcement learning settings, \citet{calvano2020artificial} demonstrate that Q-learning algorithms converge to collusive outcomes through reward-punishment strategies, and \citet{klein2021autonomous} extends this finding to sequential pricing. \citet{banchio2022artificial} provide theoretical foundations by showing how independent algorithms can develop correlated beliefs that sustain collusion.

Empirical evidence increasingly supports these theoretical findings. \citet{chen2016empirical} identify more than 500 algorithmic sellers on Amazon Marketplace, documenting that algorithmic sellers are more strategically successful and that approximately 2.4\% of all sellers employ pricing algorithms. \citet{assad2024algorithmic} provide causal evidence that algorithmic pricing adoption increased margins by 28\% in German retail gasoline duopolies. \citet{musolff2022algorithmic} shows that repricing algorithms on Amazon employ strategic price-raising mechanisms to induce competitors to follow suit, with market prices increasing by 11.4\% in concentrated markets.
 \citet{wang2023algorithms} show that simple rule-based pricing can sometimes sustain higher prices than competition among reinforcement-learning pricing agents, suggesting that seemingly unsophisticated automation may also soften price competition in markets with interacting algorithms.

Theoretically, \citet{brown2023competition} demonstrate that simple pricing algorithms can increase equilibrium prices, while \citet{abada2023artificial} show that collusion may arise from imperfect exploration in dynamic settings. \citet{meylahn2022learning} prove that collusion can emerge through gradient-based algorithms, providing the first theoretical convergence guarantees for tacit algorithmic collusion. \citet{loots2023data} construct an explicit pricing algorithm that converges to collusive prices under self-play by applying axiomatic bargaining theory, demonstrating that algorithmic collusion can emerge when firms deploy identical algorithms that infer private demand information from observable price paths.

At the same time, collusion is neither universal nor inevitable. \citet{cooper2015learning} show that learning outcomes range from Nash equilibrium to collusion depending on model specification and initial conditions. \citet{li2025adaptive} establish that when firms use individual data-driven learning algorithms without access to competitor information, they naturally converge to the Nash equilibrium, indicating that common data is necessary for collusion. \citet{li2024lego} propose a learning policy for sequential price competition that converges to the Nash equilibrium at a rate $O(1/\sqrt{T})$ without requiring inter-seller communication or coordinated price experiments. \citet{vandegeer2019dynamic} document through a dynamic pricing challenge that algorithmic performance varies substantially across market environments, confirming the complexity of pricing and learning under competition. In the related domain of assortment optimization, \citet{li2025online} develop an online learning algorithm under the Markov chain choice model with optimal regret guarantees. On the regulatory front, \citet{hartline2024regulation} develop an auditing framework for pricing algorithms based on calibrated regret, providing tools to detect tacit coordination from observed deployment data. \citet{yang2024fairness} study fairness regulation of prices in competitive markets, and \citet{yang2023regulating} examine how price fairness regulation affects the sustainability of tacit collusion, showing that strict fairness requirements can either weaken or inadvertently support collusive behavior depending on product differentiation.
Our work differs from this algorithmic pricing literature in a fundamental way. Traditional algorithms require extensive training over hundreds of thousands of periods to discover collusive strategies, whereas LLMs arrive pretrained with business knowledge and pricing practices already encoded. Our mechanism operates via a shared pricing propensity rather than independently learned strategies, a pattern related to the monoculture literature above.

{\bf LLM-based pricing and strategic behavior.}
Recent work examines LLM behavior in economic settings. \citet{fish2024algorithmic} provide the first evidence that LLM-based pricing agents converge to supra-competitive prices, with GPT-4 reaching collusive outcomes within 100 periods compared to hundreds of thousands for Q-learning. \citet{wu2024shall} corroborate this finding in Bertrand competition simulations, showing that LLMs collude within 400 rounds without explicit instructions to coordinate and form price agreements within 30 rounds when communication is permitted.
\citet{robinson2025framing} find that contextual framing strongly affects LLM cooperation: real-world business scenarios yield 98\% cooperation rates, compared with 37\% for abstract framings. This suggests that LLMs deployed for actual pricing may naturally lead to collusive outcomes.
\citet{keppo2026fragility} show that LLM collusion is fragile under agent heterogeneity: differences in patience, data access, number of competitors, and cross-algorithm diversity all reduce supra-competitive pricing, though model-size differences alone do not break collusion.
To our knowledge, our paper is the first theoretical framework to analyze how a shared AI model's pricing propensity, as exemplified by the concentrated LLM market, can produce supracompetitive pricing through the interaction of correlated recommendations and performance-driven model updating.

The collusive force we identify operates through a different mechanism from the one documented by \citet{fish2024algorithmic}. In their experiments, each LLM agent is a separate instance that receives, via a prompt, both sellers' price histories and the prompting seller's profit history, without retraining the model. Collusion arises through in-context strategic reasoning, whereby agents observe rivals' price responses and develop price-war-avoidance behavior. In our model, cross-seller information is incorporated only through data aggregation during periodic retraining. Supracompetitive pricing arises not from strategic reasoning within a single context window but from correlated recommendations driven by a shared propensity for high prices and self-reinforcing feedback loops. These two mechanisms are complementary, suggesting that the risks of supracompetitive pricing from LLM-based pricing may be broader than those indicated by either mechanism alone.

\section{Model}

{\bf The pricing game.}
We consider a symmetric duopoly pricing game in which two identical sellers compete by simultaneously choosing prices for their horizontally differentiated products over an infinite horizon (see, e.g., \citealt{banchio2022artificial}). In each round, each seller can select either a high-price strategy ($H$) or a low-price strategy ($L$), with payoffs determined by the payoff matrix in Table \ref{tab:payoff_matrix}, where $r \in (1, 2)$ denotes the relative profitability of the high-price strategy. Specifically, a higher value of $r$ makes the high-price decision more attractive.
Since $r < 2$ implies $2+r > 2r$ and $r > 1$ implies $2 > r$, the strategy $L$ strictly dominates $H$, making $(L, L)$ the unique Nash equilibrium with a competitive payoff of $2$ per seller. The Pareto-superior outcome $(H, H)$, yielding $2r > 2$ per seller, serves as the collusive benchmark.

\begin{table}[h]
\centering
\begin{tabular}{cc|c|c|}
& \multicolumn{1}{c}{} & \multicolumn{2}{c}{Seller 2}\\
& \multicolumn{1}{c}{} & \multicolumn{1}{c}{$H$}  & \multicolumn{1}{c}{$L$} \\\cline{3-4}
\multirow{2}*{Seller 1}  & $H$ & $(2r,2r)$ & $(r,2+r)$ \\\cline{3-4}
& $L$ & $(2+r,r)$ & $(2,2)$ \\\cline{3-4}
\end{tabular}
\caption{Profit structure of the duopolistic price competition, where $1 < r < 2$. $H$ represents the high-price strategy, while $L$ indicates the low-price strategy. Adapted from \citet{banchio2022artificial}.}
\label{tab:payoff_matrix}
\end{table}

{\bf Shared AI model.}
We study a stylized model of shared AI-based pricing. While LLMs are the leading motivating example, given their market concentration and data-sharing practices documented above, our framework applies more broadly to any setting in which competing sellers receive pricing recommendations from a common AI model with two features: (i) a shared pricing propensity, encoded in the model's weights, that creates cross-seller correlation, and (ii) periodic model updating that aggregates performance feedback from multiple sellers. We deliberately abstract from firm-specific prompt histories, in-context strategic reasoning, and heterogeneous model access. These features are important in practice but orthogonal to the monoculture mechanism we isolate. We discuss them as future research directions in the Section~\ref{sec:conclusion}.

Both sellers delegate their pricing decisions to a shared AI model by requesting pricing recommendations. We use the term ``shared AI model'' to encompass settings ranging from both sellers querying the same LLM provider to using different providers whose models share common training data, architecture, or alignment procedures. In all cases, the key feature is that the model's weights, and therefore its learned pricing propensity, are common across sellers. We assume that sellers adopt these recommendations rather than merely soliciting them. This assumption reflects the operational reality of algorithmic decision-making: research on algorithm appreciation shows that decision-makers systematically follow algorithmic advice, often preferring it over human judgment \citep{logg2019algorithm}. Once sellers delegate pricing to an AI model, manual override becomes costly. Such interventions require dedicated analyst time, disrupt automated workflows, and introduce inconsistency that complicates performance attribution. Moreover, the value proposition of LLM-based pricing lies in enabling rapid, scalable, data-driven decisions. This advantage erodes when human judgment routinely second-guesses each recommendation.

While we assume full adoption for expositional clarity, this assumption can be partially relaxed within our framework. If sellers instead adopt recommendations with probability less than 1, this partial adherence introduces additional randomness that reduces the effective correlation between the model's pricing propensity and the sellers' final adopted pricing decision. Our framework fully captures this case via the output-fidelity parameter defined below, though it should be interpreted more broadly than its nominal meaning.

The model's pricing behavior is characterized by two fundamental parameters: the \emph{propensity parameter} $\theta \in [0,1]$ and the \emph{output fidelity} $\rho \in [0.5,1]$.
The former parameter $\theta$ represents the proportion of the model's internal recommendation policy that favors high-price outputs. Formally, the model maintains a latent preference state unobservable to sellers, operating in either a high-price mode (denoted $H$-mode) or a low-price mode (denoted $L$-mode).
This parameter is a reduced-form representation of the model's pricing propensity: its structural tendency toward certain pricing strategies, shaped by training data, model architecture, and any alignment or fine-tuning procedures.
Because both sellers rely on a common AI model, this latent preference, and therefore the model's pricing propensity, is shared between sellers, creating an endogenous correlation in their pricing recommendations. The propensity parameter $\theta$ remains fixed until the model is retrained, at which point it is updated based on observed performance outcomes from seller interactions.

When $\theta > 1/2$, the $H$-mode is more likely, so the model tends toward high-price recommendations. Conversely, when $\theta < 1/2$, the $L$-mode predominates, and low prices are favored. At the symmetric point $\theta = 1/2$, the model is equally likely to operate in either mode, representing maximal uncertainty in its pricing preference.

The output fidelity $\rho$ quantifies the alignment between the model's pricing propensity and the pricing decisions sellers actually adopt. Multiple factors contribute to imperfect alignment: stochastic generation mechanisms (e.g., sampling temperature and top-$k$/top-$p$ filtering), differences in seller-specific prompts or contexts, and partial seller adherence to recommendations. In our reduced-form model, $\rho$ aggregates all these sources of noise into a single parameter. The model does not deterministically produce recommendations that perfectly align with its latent preference. Specifically, conditional on adopting the $H$-mode, the model recommends strategy $H$ to each seller with probability $\rho$ and strategy $L$ with probability $1-\rho$. Conversely, conditional on the $L$-mode, the model recommends $L$ with probability $\rho$ and $H$ with probability $1-\rho$. It is worth noting that these output realizations are conditionally independent across the two sellers, given the shared latent preference, introducing idiosyncratic noise into each seller's recommendation.

When $\rho \to 1$, the shared latent preference dominates, inducing strong positive correlation in seller strategies. In contrast, when $\rho \to 0.5$, independent output noise overwhelms the preference signal, rendering strategies increasingly randomized despite the shared underlying model with a common propensity parameter $\theta$.

This formulation also accommodates partial seller adherence, as previewed above. If a seller overrides the model's recommendation with some probability, independently choosing a pricing strategy, this additional randomness reduces the effective alignment between the model's latent preference and the seller's realized action. Formally, partial adherence is equivalent to lowering the effective value of $\rho$. The output fidelity parameter thus captures both the model's generation stochasticity and the degree of seller adherence, with higher adherence corresponding to higher effective $\rho$.

Given the parameters $(\theta, \rho)$, we can then derive the probability distribution over strategy profiles. Let $p_{ij}$ denote the probability that Seller 1 chooses strategy $i$ and Seller 2 chooses strategy $j$, where $i,j \in \{H,L\}$. The joint probabilities are:
\begin{align}
p_{HH}(\theta) &= \theta \rho^2 + (1-\theta) (1-\rho)^2, \label{eq:pHH}\\
p_{HL}(\theta) = p_{LH}(\theta) &= \rho(1-\rho), \label{eq:pHL}\\
p_{LL}(\theta) &= \theta (1-\rho)^2 + (1-\theta) \rho^2. \label{eq:pLL}
\end{align}
Equation \eqref{eq:pHH} decomposes as follows: $\theta\rho^2$ represents the probability that the model favors high prices and both outputs align with this preference, while $(1-\theta)(1-\rho)^2$ captures cases where the model favors low prices but both outputs deviate, yielding $H$ for both sellers. Similar interpretations apply to \eqref{eq:pHL} and \eqref{eq:pLL}.

{\bf Preference update mechanism.}
The shared AI model updates its propensity parameter $\theta$ through periodic retraining or fine-tuning on seller interactions. Specifically, the model adjusts its pricing propensity based on the relative performance of high-price versus low-price recommendations. Intuitively, if high-price recommendations yield higher average payoffs than low-price recommendations, the model should increase its propensity toward high prices, and vice versa.

This update mechanism requires the model to observe, or infer, both sellers' pricing actions and realized payoffs.
Such observability may arise through several channels: explicit data sharing between sellers and the AI provider, user satisfaction feedback (e.g., thumbs-up/down ratings), behavioral signals such as recommendation adoption or abandonment, and publicly observable pricing actions on comparison platforms. Our stochastic approximation framework requires only that the feedback signal be an unbiased (or asymptotically unbiased) estimator with bounded variance, and it accommodates noisy or incomplete outcome observation. This parallels standard assumptions in the algorithmic collusion literature (see, e.g., \citealt{calvano2020artificial,banchio2022artificial}).

To formalize this intuition, we consider a batch learning framework in which the model is retrained after observing every batch of $b \geq 1$ decision rounds. At each retraining step, the model evaluates the relative performance of strategies $H$ and $L$ from the observed data and updates its propensity accordingly. We then consider an update rule that correctly tracks the conditional payoff difference between strategies, accounting for the dependence of strategy frequencies on the current propensity.

Let $S \in \{H, L\}$ denote a randomly sampled seller's action and $\Pi$ denote the seller's realized profit in a given round. Define the \emph{conditional payoff difference}
\begin{equation}\label{eq:Delta-def}
\Delta(\theta) = \mathbb{E}[\Pi \mid S=H, \theta] - \mathbb{E}[\Pi \mid S=L, \theta],
\end{equation}
which measures the expected payoff advantage of the high-price strategy over the low-price strategy, conditional on being recommended. The sign of $\Delta(\theta)$ determines which strategy yields higher expected returns under the current propensity.
A natural learning objective is for the model to increase $\theta$ when $\Delta(\theta) > 0$ (the high price outperforms) and decrease $\theta$ when $\Delta(\theta) < 0$ (the low price outperforms). However, since $\Delta(\theta)$ is not directly observable, the model must estimate it from finite samples of interaction data.

{\bf Estimating $\Delta(\theta)$ from sellers' feedback.} To update its propensity, the model must estimate the conditional payoff difference $\Delta(\theta)$ from observed interaction data. A fundamental challenge arises: the model observes payoffs only for strategies that are actually recommended and adopted, not for the counterfactual outcomes under alternative recommendations. To capture this selection structure, we model the estimation procedure using inverse probability weighting (IPW), a classical causal inference technique that corrects for selection bias by reweighting observations based on their sampling probabilities. We emphasize that our theoretical framework extends beyond this particular specification: the critical requirement is that the estimator be unbiased with bounded variance. Any estimation procedure satisfying these mild conditions will asymptotically track the same underlying performance measure, as formally established in our subsequent analysis. In particular, IPW serves as a canonical representative within this broader class, enabling a clean characterization of the model's learning dynamics while connecting naturally to the established literature on policy evaluation (see, e.g., \citealp{horvitz1952generalization,imbens2015causal}).

Consider a single round of interaction producing actions $(S_1, S_2)$ and payoffs $(\Pi_1, \Pi_2)$ for the two sellers. For each seller $i \in \{1,2\}$ we observe the pair $(S, \Pi) = (S_i, \Pi_i)$.
Let $$
p_H(\theta)\;\triangleq\; \mathbb{P}(S=H\mid \theta) = \theta\rho+(1-\theta)(1-\rho)$$
denote the marginal probability that a seller receives recommendation $H$ when the propensity is $\theta$, and let $p_L(\theta) = 1 - p_H(\theta)$. The key observation is that the observed strategy $S$ is drawn from a known distribution that depends on $\theta$, enabling inverse probability weighting.
In particular, we define the estimator
\begin{equation}\label{eq:D-def}
D(\theta) = \frac{\Pi}{p_H(\theta)}\,\mathbbm{1}\{S=H\} - \frac{\Pi}{p_L(\theta)}\,\mathbbm{1}\{S=L\}.
\end{equation}
This construction provides an unbiased estimate of the conditional payoff difference from a single observation.

\begin{lemma}[{\sc Unbiased Estimation}]\label{lem:unbiased-D}
For any fixed $\theta \in [0,1]$ with $p_H(\theta), p_L(\theta) \in (0,1)$, the random variable $D(\theta)$ defined in \eqref{eq:D-def} satisfies
\[
\mathbb{E}[D(\theta)\mid \theta]=\Delta(\theta).
\]
Moreover, for fixed $\rho\in(1/2,1)$, $D(\theta)$ is uniformly bounded over $\theta\in[0,1]$.
\end{lemma}

{\bf A principled learning rule.}
Given the unbiased estimator, we now derive an update rule for the model's propensity. At the $n$th retraining step, the model observes a batch of $b$ seller feedback rounds. In particular, for each round $\ell\in\{1,\dots,b\}$ and seller $i\in\{1,2\}$, we observe $(S_i^{(\ell)},\Pi_i^{(\ell)})$ and compute $D_i^{(\ell)}(\theta_n)$ via \eqref{eq:D-def}. The batch-average estimator is
\begin{equation}\label{eq:batch-average}
\overline{D}_n^{(b)} = \frac{1}{2b}\sum_{\ell=1}^b \sum_{i=1}^2 D_i^{(\ell)}(\theta_n).
\end{equation}
This batch average remains an unbiased estimator $\mathbb{E}[\overline{D}_n^{(b)} \mid \theta_n] = \Delta(\theta_n)$.

A standard and theoretically grounded way to update a Bernoulli propensity is to update the \emph{log-odds}. In the learning literature (see, e.g., \citealp{cesabianchi2006prediction}), it is equivalent to applying exponential weights or mirror descent on the two-action simplex. Specifically, let
\[
z_n \;\triangleq\; \log\!\Bigl(\frac{\theta_n}{1-\theta_n}\Bigr)\in\mathbb{R},
\qquad
\theta_n \;=\; \sigma(z_n)\;\triangleq\; \frac{1}{1+e^{-z_n}}.
\]
Given step sizes $(\gamma_n)_{n\ge 0}$ with $\gamma_n > 0$, $\sum_{n=0}^\infty \gamma_n = \infty$, and $\sum_{n=0}^\infty \gamma_n^2 < \infty$ (for example, $\gamma_n = \eta/(n+1)^\alpha$ with $\alpha\in(1/2,1]$ and $\eta\in(0,1]$), the model's propensity evolves according to
\begin{equation}\label{eq:logit-recursion}
z_{n+1} \;=\; z_n \;+\; \gamma_n\,\overline{D}_n^{(b)},\qquad \theta_{n+1}=\sigma(z_{n+1}).
\end{equation}

This update rule employs batch computations for iterative parameter refinement, naturally aligning with the stochastic gradient descent procedures used in LLM training. Moreover, it admits an intuitive interpretation.
We note that
\[
\theta_{n+1}-\theta_n
= \sigma(z_n+\gamma_n\overline{D}_n^{(b)})-\sigma(z_n)
= \sigma'(z_n)\gamma_n\overline{D}_n^{(b)}+O(\gamma_n^2),
\]
and $\sigma'(z)=\sigma(z)(1-\sigma(z))=\theta_n(1-\theta_n)$. Hence,
\begin{equation}\label{eq:theta-firstorder}
\theta_{n+1} \;=\; \theta_n \;+\; \gamma_n\,\theta_n(1-\theta_n)\,\overline{D}_n^{(b)} \;+\; O(\gamma_n^2).
\end{equation}
Equation~\eqref{eq:theta-firstorder} is the first-order approximation of the log-odds update and corresponds to a two-action replicator step in evolutionary game theory \citep{hofbauer1998evolutionary}, or a natural-gradient step from the perspective of information geometry \citep{amari1998natural}. Intuitively, the model adjusts its propensity in the direction indicated by the payoff difference estimator. The scaling factor $\theta_n(1-\theta_n)$ ensures that updates are larger when the propensity is intermediate and the uncertainty about which strategy is better is high, and updates are smaller near the boundaries $\theta = 0$ or $\theta = 1$, where the model already has strong beliefs.

{\bf Connection to policy optimization.}
The log-odds update rule \eqref{eq:logit-recursion} admits a natural optimization interpretation. Consider the expected payoff of a randomly sampled seller under propensity $\theta$:
\[
V(\theta) = p_H(\theta)\,\mathbb{E}[\Pi \mid S=H, \theta] + p_L(\theta)\,\mathbb{E}[\Pi \mid S=L, \theta].
\]
The gradient of $V$ with respect to the log-odds $z = \log(\theta/(1-\theta))$ can be shown to equal $\theta(1-\theta)\Delta(\theta)$, which is precisely the drift term in Equation~\eqref{eq:theta-firstorder}. The update rule \eqref{eq:logit-recursion} thus performs stochastic natural-gradient ascent (or equivalently, mirror descent with the negative-entropy regularizer on the two-action simplex) on the expected payoff objective $V(\theta)$. This connection ensures that the learning dynamics are not ad hoc but correspond to a principled optimization of recommendation quality from the model's perspective.

\begin{definition}[{\sc Supracompetitive vs.\ Competitive Pricing}]\label{def:outcomes}
The long-run market outcome exhibits \emph{supracompetitive pricing} (also referred to as \emph{collusive pricing} when connecting to the algorithmic collusion literature) if the model's propensity $\theta$ converges to a value $\theta^* > 0$ such that the joint high-pricing probability $p_{HH}(\theta^*)$ and the expected per-seller payoff both exceed their competitive-equilibrium levels (i.e., the levels arising when $\theta = 0$). The outcome exhibits \emph{competitive pricing} if the model converges to $\theta = 0$, so that $p_{HH} \to (1-\rho)^2$ and expected payoffs approach the Nash equilibrium level of $2$ per seller. We emphasize that supracompetitive pricing in our framework describes a market outcome, not a strategic process: it does not presuppose the punishment strategies, discount factors, or explicit coordination that characterize collusion in classical repeated-game models.
\end{definition}

\section{Analysis}

We analyze the model's long-run behavior in making pricing recommendations under the update rule \eqref{eq:logit-recursion}. In particular, we first study the large-batch regime ($b\to\infty$), in which stochastic noise vanishes. We then consider the finite-batch case using stochastic approximation, and analyze the effect of batch size.
Before proceeding, we first note that the stochastic recursion \eqref{eq:logit-recursion} admits a deterministic characterization in terms of an associated ordinary differential equation (ODE). The following lemma, which adapts classical results from stochastic approximation theory \citep{borkar2008,benaim1999}, formalizes this connection.

\begin{lemma}[{\sc ODE Tracking}]\label{lem:ODE-tracking}
Consider the stochastic recursion \eqref{eq:logit-recursion} and let $(\mathcal{H}_n)_{n \geq 0}$ denote the natural filtration generated by the sequence $(z_0, z_1, \ldots, z_n)$. Suppose the following conditions hold:
\begin{enumerate}
\item[\emph{(A1)}] \emph{step-size conditions:} The learning rates satisfy $\gamma_n > 0$ for all $n$, $\sum_{n=0}^\infty \gamma_n = \infty$, and $\sum_{n=0}^\infty \gamma_n^2 < \infty$.
\item[\emph{(A2)}] \emph{unbiased gradient estimation:} The batch-average estimator satisfies $\mathbb{E}[\overline{D}_n^{(b)} \mid \mathcal{H}_n] = \Delta(\theta_n)$.
\item[\emph{(A3)}] \emph{bounded second moments:} For any fixed $\rho \in (1/2, 1)$, there exists a constant $C < \infty$ such that $\mathbb{E}[(\overline{D}_n^{(b)})^2] \leq C$ for all $n$.
\end{enumerate}
Then we have
\begin{enumerate}
\item[\emph{(i)}] \emph{(Asymptotic Tracking)} Trajectory $(z_0, z_1, \dots, z_n)$ asymptotically tracks the solution to the mean-field ODE
\begin{equation}\label{eq:ODE-z}
\dot{z}(t) = \Delta\bigl(\sigma(z(t))\bigr).
\end{equation}
Formally, define the continuous-time interpolation $\bar{z}:[0,\infty) \to \mathbb{R}$ by setting $t_0 = 0$, $t_n = \sum_{k=0}^{n-1} \gamma_k$ for $n \geq 1$, and letting $\bar{z}(t)$ be the piecewise-linear function satisfying $\bar{z}(t_n) = z_n$. For any initial condition $z \in \mathbb{R}$, let $\Phi_t(z)$ denote the solution to the ODE \eqref{eq:ODE-z} at time $t$ when starting from $z(0) = z$. Then for any $T > 0$,
\[
\sup_{0 \leq t \leq T} \bigl|\bar{z}(t_n + t) - \Phi_t(z_n)\bigr| \xrightarrow{n \to \infty} 0 \quad \text{almost surely}.
\]
\item[\emph{(ii)}] \emph{(Equilibrium Convergence)} Almost surely, every limit point of the sequence $(z_n)_{n \geq 0}$ is an equilibrium of the ODE \eqref{eq:ODE-z}, that is, a point $z^\star$ satisfying $\Delta(\sigma(z^\star)) = 0$.
\end{enumerate}
\end{lemma}

Lemma~\ref{lem:ODE-tracking} provides a theoretical foundation for analyzing the long-run behavior of LLM-based pricing. The result states that the noisy, discrete-time learning process \eqref{eq:logit-recursion} behaves like the smooth, deterministic dynamical system \eqref{eq:ODE-z} over long time horizons. This ``ODE method'' reduces the analysis of a complex stochastic system to the study of a one-dimensional differential equation, whose equilibria and stability properties can be characterized explicitly. We will exploit this connection throughout our analysis.

Intuitively, condition (A1) ensures that the step sizes $\gamma_n$ decay slowly enough for the algorithm to explore the parameter space (via $\sum_{n=0}^\infty \gamma_n = \infty$), yet fast enough for the noise to average out (via $\sum_{n=0}^\infty \gamma_n^2 < \infty$). Condition (A2) guarantees that the estimator $\overline{D}_n^{(b)}$ provides an unbiased signal for the true payoff advantage $\Delta(\theta_n)$, so that on average the updates point in the ``correct'' direction. Condition (A3) bounds the estimation variance, preventing the noise from overwhelming the signal.
Under these conditions, the cumulative effect of the martingale noise terms vanishes in the limit, leaving only the deterministic drift $\Delta(\theta)$. The sequence $(z_n)_{n\ge0}$ thus tracks the integral curves of the ODE, and its limit points must satisfy the equilibrium condition $\Delta(\sigma(z^\star)) = 0$.

We next provide an equivalent formulation of \eqref {eq:ODE-z} in propensity space.
Specifically, applying the chain rule to the transformation $\theta(t) = \sigma(z(t))$, where $\sigma'(z) = \sigma(z)(1-\sigma(z))$, yields an equivalent ODE
\begin{equation}\label{eq:ODE}
\dot{\theta}(t) = \theta(t)\bigl(1-\theta(t)\bigr)\,\Delta(\theta(t)).
\end{equation}
This characterization will prove convenient for characterizing equilibria and their stability properties. Intuitively, the factor $\theta(1-\theta)$ ensures that the boundaries $\theta \in \{0,1\}$ are absorbing, while the sign of $\Delta(\theta)$ determines whether the propensity drifts toward high-price or low-price recommendations.

We characterize the equilibrium structure of the dynamical system \eqref{eq:ODE} by first analyzing the asymptotic regime where the batch size $b \to \infty$. This limiting case provides analytical tractability while capturing the essential equilibrium dynamics.

\subsection{Large-Batch Regime}
\label{sec:large-b-limit}

When batch size $b$ grows large, the law of large numbers implies that the batch-averaged estimator $\overline{D}_n^{(b)}$ concentrates around its expectation $\Delta(\theta_n)$. As $b \to \infty$, the stochastic recursion \eqref{eq:logit-recursion} converges almost surely to a deterministic recursion that tracks the ODE \eqref{eq:ODE}
\begin{equation}\label{eq:deterministic-update-z}
z_{n+1} = z_n + \gamma_n \Delta(\theta_n),\qquad \theta_n=\sigma(z_n).
\end{equation}
Consequently, the model increases its propensity toward a high-price strategy almost surely when $\Delta(\theta_n) > 0$, that is, for all $\theta_n = \theta$ such that
\begin{equation*}
\frac{p_{HH}(\theta) \cdot 2r + p_{HL}(\theta) \cdot r}{p_{HH}(\theta) + p_{HL}(\theta)} > \frac{p_{LH}(\theta) \cdot (r+2) + p_{LL}(\theta) \cdot 2}{p_{LH}(\theta) + p_{LL}(\theta)}.
\end{equation*}
The left-hand side represents the expected profit conditional on receiving recommendation $H$, while the right-hand side is the expected profit conditional on recommendation $L$. The following proposition characterizes the region in $(\theta,\rho)$ space where this inequality holds.

\begin{proposition}[{\sc Conditions for Price Propensity Drift}]\label{prop:H-dominance}
Let $r\in(1,2)$ and $\rho\in(1/2,1]$. Define
\begin{equation}\label{eq:s-definition}
s(\rho,r) \;\triangleq\; \frac{2(2-r)\,\rho(1-\rho)}{(r-1)(2\rho-1)^2},
\end{equation}
and when $s(\rho,r)\le 1$ define
\begin{equation}\label{eq:theta-thresholds}
\theta_\pm(\rho,r) \;\triangleq\; \frac{1}{2} \pm \frac{1}{2}\sqrt{1-s(\rho,r)}.
\end{equation}
Then
\begin{enumerate}[(i)]
\item If $s(\rho,r) > 1$, then $\Delta(\theta) < 0$ for all $\theta \in (0,1)$.
\item If $s(\rho,r) = 1$, then $\Delta(\theta)=0$ if and only if $\theta=1/2$, and $\Delta(\theta)<0$ for all $\theta\in(0,1)\setminus\{1/2\}$.
\item If $s(\rho,r) < 1$, then $\Delta(\theta) > 0$ if and only if $\theta \in (\theta_-(\rho,r), \theta_+(\rho,r))$, with $\Delta(\theta_\pm)=0$ and $\Delta(\theta)<0$ outside $[\theta_-,\theta_+]$.
\end{enumerate}
\end{proposition}

Proposition~\ref{prop:H-dominance} characterizes when the learning dynamics almost surely push the model's propensity $\theta$ upward, toward more frequent high-price recommendations. The sign of the conditional payoff difference $\Delta(\theta)$ governs this direction: when $\Delta(\theta)>0$, the high-price recommendation $H$ yields higher expected profit conditional on being played, reinforcing the model's propensity for $H$. The proposition partitions the parameter space into regions determined by the statistic $s(\rho,r)$, which captures the ratio of two competing forces. In particular, the numerator in \eqref{eq:s-definition} represents the cost of recommendation deviations. When one seller receives $H$ and the other receives $L$, the low-price seller earns $2+r$ while the high-price seller earns only $2r$, leading to a gap of $2-r$ favoring $L$. The factor $\rho(1-\rho)$ is the probability of such a deviation per seller pair.
On the other hand, the denominator in \eqref{eq:s-definition} implies the benefit of coordinated high pricing. When both sellers receive $H$ and play $(H,H)$, each earns $2r$ rather than the competitive payoff of $2$; the per-seller gain is $2(r-1)$. The factor $(2\rho-1)^2$ measures the correlation strength in recommendations induced by the shared latent mode.
When $s(\rho,r)<1$, the coordination benefit dominates the deviation cost, creating a parameter region where high-price recommendations are self-reinforcing.

For any fixed $r\in(1,2)$, the statistic $s(\rho,r)$ is strictly decreasing in $\rho$ over $(1/2,1)$: it diverges to $+\infty$ as $\rho\to 1/2^+$ and vanishes as $\rho\to 1$. This monotonicity reflects a fundamental tradeoff: higher fidelity simultaneously reduces miscoordinations (i.e., $\rho(1-\rho)$ decreases) and strengthens coordination (i.e., $(2\rho-1)^2$ increases). Consequently, for any profitability parameter $r\in(1,2)$, there exists a critical threshold $\rho_c(r)\in(1/2,1)$ above which $s(\rho,r)<1$ and an upward-drift region for price propensity emerges. Therefore, sufficiently high output fidelity always makes supracompetitive dynamics possible.

We also note that even when $s(\rho,r)<1$, the region where $\Delta(\theta)>0$ is an interval $(\theta_-,\theta_+)$ bounded away from both $0$ and $1$. The intuition is that at extreme propensities, the uncommon recommendation almost always signals a deviation from the recommendation, to the opponent, and deviations always favor the low-price seller.
Specifically, when $\theta$ is near zero and propensity strongly favors $L$, the latent mode is almost always $L$. A seller who nonetheless receives $H$ is likely an outlier, while the opponent almost surely receives $L$. The resulting $(H,L)$ outcome punishes the $H$-playing outlier, who earns only $r$ compared to the opponent's $2+r$. Hence $\mathbb{E}[\Pi\mid S=H]\approx r < 2 \approx \mathbb{E}[\Pi\mid S=L]$, and learning drives $\theta$ further toward zero.
In contrast, when $\theta$ is near one and propensity strongly favors $H$, the mode is almost always $H$. A seller who receives $L$ is the rare outlier, likely facing an opponent who receives $H$. The $(L,H)$ outcome rewards this outlier with $2+r$, exceeding the $(H,H)$ payoff of $2r$ (since $r<2$). Hence $\mathbb{E}[\Pi\mid S=L]\approx 2+r > 2r \approx \mathbb{E}[\Pi\mid S=H]$, and learning pushes $\theta$ back down.
These asymmetric outcomes drive $\Delta(\theta)<0$ at both extremes: near $\theta=0$ the rare $H$-recommendation is punished, while near $\theta=1$ the rare $L$-recommendation is rewarded, but this makes $L$ preferable to $H$. This creates a \emph{basin of attraction} (defined as the region of those starting points that converge to a particular equilibrium) around the fixed point $\theta_+$ when it exists, while preserving $\theta=0$ as a competing attractor.

\begin{proposition}[{\sc Collusive Behavior in the Large-Batch Limit}]\label{prop:convergence-binf}
Fix $r \in (1,2)$, $\rho \in (1/2, 1]$, a step-size exponent $\alpha\in(1/2,1]$, and a scaling constant $\eta\in(0,1]$. Consider the step sizes
\(
\gamma_n = \frac{\eta}{(n+1)^\alpha},
\)
so that $\sum_{n=0}^\infty \gamma_n = \infty$ and $\sum_{n=0}^\infty \gamma_n^2 < \infty$.
Let $(\theta_n^{(b)})_{n \geq 0}$ denote the sequence generated by the stochastic recursion \eqref{eq:logit-recursion} with batch size $b$ and initial condition $\theta_0 \in (0,1)$. Define the critical output-fidelity threshold
\begin{equation}\label{eq:rho-critical}
\rho_c(r) \;\triangleq\; \frac{1 + \sqrt{(2-r)/r}}{2}.
\end{equation}
If $\rho\in(1/2,1)$, then for any fixed $N<\infty$, as $b \to \infty$ we have
\[
\max_{0\le n\le N}\bigl|\theta_n^{(b)}-\theta_n\bigr|\xrightarrow{\emph{a.s.}} 0,
\]
where $(\theta_n)_{n \geq 0}$ is the solution of the limiting recursion \eqref{eq:deterministic-update-z} with the same initial condition.
Moreover, as $n \to \infty$, the limiting recursion satisfies:
\begin{enumerate}[(i)]
\item \emph{(Low-fidelity regime)} If $\rho \in (1/2, \rho_c(r))$, then $\theta_n \to 0$ for all $\theta_0 \in (0,1)$.
\item \emph{(Critical regime)} If $\rho = \rho_c(r)$, then $\theta_n \to 0$ for $\theta_0 \in (0, 1/2)$, and $\theta_n \to 1/2$ for $\theta_0 \in [1/2, 1)$.
\item \emph{(High-fidelity regime)} If $\rho \in (\rho_c(r), 1)$, then:
\begin{enumerate}
\item[(a)] $\theta_n \to 0$ if $\theta_0 \in (0, \theta_-(\rho,r))$;
\item[(b)] $\theta_n \to \theta_+(\rho,r) \in (1/2, 1)$ if $\theta_0 \in (\theta_-(\rho,r), 1)$;
\item[(c)] $\theta_n \to \theta_-(\rho,r)$ if $\theta_0 = \theta_-(\rho,r)$ (unstable equilibrium).
\end{enumerate}
\item \emph{(Perfect-fidelity regime)} If $\rho = 1$, then $\theta_n \to 1$ for all $\theta_0 \in (0,1)$.
\end{enumerate}
\end{proposition}

Proposition~\ref{prop:convergence-binf} characterizes the long-run behavior of LLM-based pricing when retraining occurs over sufficiently large batches of interaction data. The analysis builds on Lemma~\ref{lem:ODE-tracking}: as $b \to \infty$, the stochastic recursion converges to its deterministic limit, and the ODE tracking result ensures that long-run outcomes correspond to equilibria of the mean-field ODE. In this regime, estimation noise becomes negligible, and the learning dynamics are governed entirely by the sign of the conditional payoff difference $\Delta(\theta)$. The proposition reveals the \emph{emergence of supracompetitive pricing} through a phase transition in the system's long-run behavior, governed by the critical threshold $\rho_c(r)$. This threshold marks the minimum output fidelity above which a collusive equilibrium exists: when $\rho > \rho_c(r)$, the learning dynamics can sustain elevated prices in the long run, whereas when $\rho < \rho_c(r)$, the unique long-run outcome is competitive pricing.

Specifically, in the \emph{low-fidelity regime} ($\rho < \rho_c(r)$), the model's recommendations are sufficiently noisy that miscoordination events where one seller prices high and the other prices low occur frequently. Because the low-price seller captures the market in such events, the expected payoff conditional on recommending $H$ falls below that of recommending $L$. The learning dynamics, therefore, push the propensity $\theta$ toward zero, regardless of the initial condition. In the long run, the model learns to recommend competitive (low) prices.

In the \emph{critical regime} ($\rho = \rho_c(r)$), the coordination benefit of mutual high pricing exactly balances the deviation cost. The equilibrium $\theta = 1/2$ exhibits one-sided stability: initial conditions $\theta_0 \geq 1/2$ converge to the symmetric equilibrium of $\theta=1/2$, while initial conditions $\theta_0 < 1/2$ drift toward competitive pricing.

In the \emph{high-fidelity regime} ($\rho_c(r) < \rho < 1$), the shared latent preference enables supracompetitive pricing when $\theta$ is sufficiently high. Specifically, when $\theta$ exceeds the threshold $\theta_-(\rho,r)$, the model frequently operates in $H$-mode. In this mode, both sellers receive the high-price recommendation with probability $\rho$, creating positive correlation in their pricing decisions. The resulting $(H,H)$ outcomes yield the collusive payoff $2r$ per seller, which exceeds the competitive payoff of $2$. This coordination benefit outweighs the cost of occasional miscoordination (in which one seller receives $H$ while the other receives $L$), so the expected payoff conditional on receiving $H$ exceeds that conditional on receiving $L$, and learning reinforces this elevated propensity. The system thus exhibits \emph{bistability}: both $\theta = 0$ (competitive pricing) and $\theta = \theta_+(\rho,r)$ (collusive pricing) are locally stable, with basins of attraction separated by the unstable threshold $\theta_-(\rho,r)$. The long-run outcome depends on the initial condition: if the model begins with a sufficiently strong prior toward low-price recommendations ($\theta_0 < \theta_-$), it converges to competitive pricing. Otherwise, it converges to the collusive equilibrium. It is worth noting that since $\theta_+(\rho,r) < 1$, this collusive outcome emerges even though the model does not recommend the high price with certainty. The equilibrium is sustained because the correlation induced by the shared latent mode generates sufficiently frequent $(H,H)$ outcomes to maintain the coordination benefit, even as occasional miscoordination events occur.

In the \emph{perfect-fidelity regime} ($\rho = 1$), recommendations are deterministic given the latent mode. Both sellers always receive identical recommendations, eliminating miscoordination entirely. Since mutual high pricing ($2r$) always exceeds the low-pricing payoff ($2$), the learning dynamics drive the propensity toward full price coordination ($\theta \to 1$) from any interior initial condition.

Proposition~\ref{prop:convergence-binf} provides a precise characterization of when LLM-based pricing leads to collusive outcomes. To better understand these collusive dynamics, we illustrate the proposition with the following example.

\begin{figure}[!htb]
    \centering
    \includegraphics[width=0.7\linewidth]{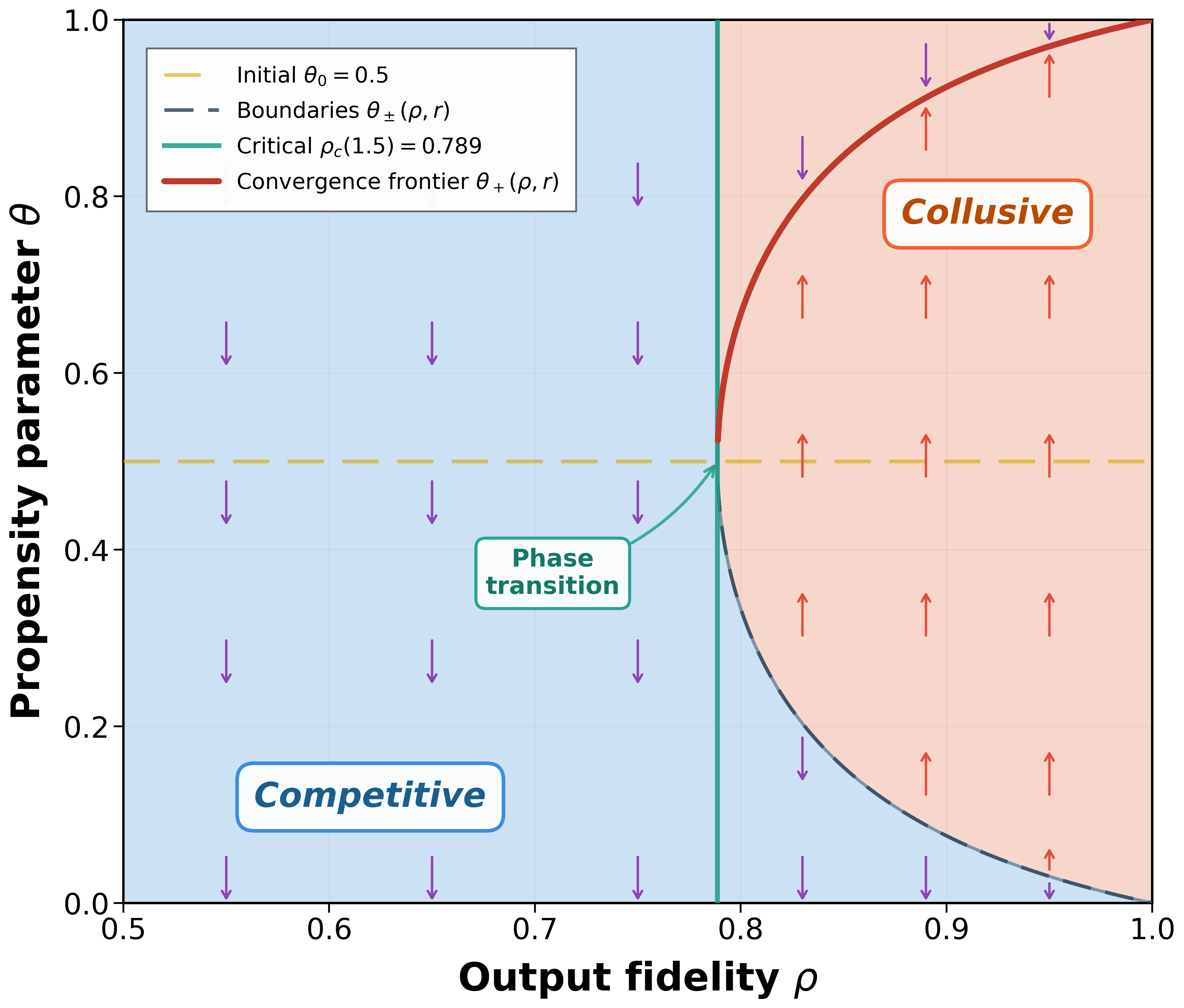}
    \caption{Phase diagram of pricing dynamics in the large batch limit ($b\to\infty$) for $r=1.5$. The shaded regions represent basins of attraction: the blue region is the competitive basin where the propensity converges to $\theta=0$, while the pink region is the collusive basin where the propensity converges to the stable equilibrium $\theta_+(\rho,r)$. The solid red curve shows the convergence frontier $\theta_+(\rho,r)$, the long-run collusive equilibrium for each $\rho\ge\rho_c(r)$. The dashed gray curves mark the boundaries $\theta_\pm(\rho,r)$: the lower boundary $\theta_-$ separates the two basins, while the region between $\theta_-$ and $\theta_+$ is where $\Delta(\theta)>0$ (high-price strategy $H$ outperforms $L$). The vertical line indicates the critical threshold $\rho_c(1.5)\approx 0.789$, separating the competitive regime (left) from the bistable regime (right). The horizontal dashed line shows the initial condition $\theta_0=1/2$. Arrows indicate the direction of propensity evolution: red upward arrows where $\Delta(\theta)>0$, purple downward arrows where $\Delta(\theta)<0$.}
    \label{fig:H-dominance-region}
\end{figure}

\begin{example}[{\sc Illustrative Visualization of Collusion Dynamics}]\label{ex:collusion-region}
We display the $(\rho,\theta)$ parameter space as stated in Proposition~\ref{prop:H-dominance} for $r=1.5$ and $\theta_0=1/2$ in Figure~\ref{fig:H-dominance-region}.
The $\Delta(\theta) > 0$ region is bounded by the curves $\theta_-(\rho,r)$ and $\theta_+(\rho,r)$, which collapse to a single point at $\theta=1/2$ when $\rho=\rho_c(r)$. For $\rho<\rho_c(r)$, no such region exists, the low-price strategy $L$ strictly dominates $H$ for all propensities $\theta\in(0,1)$, always driving the system toward competitive pricing ($\theta\to 0$).

The convergence frontier $\theta_+(\rho,r)$ illustrates the limit dynamics. Starting from the symmetric initial condition $\theta_0=1/2$ (see the horizontal dashed line), the propensity evolves according to the update rule. For any $\rho\ge\rho_c(r)$, the system converges to $\theta_+(\rho,r)$, which increases monotonically with $\rho$ and approaches 1 as $\rho\to 1$. This monotonicity implies that higher output fidelity leads to more aggressive supracompetitive pricing in the long run. For example, at $\rho = 0.85$ and $r = 1.5$, the supracompetitive equilibrium is $\theta_+ \approx 0.846$, yielding a joint high-pricing probability of $p_{HH}(\theta_+) \approx 0.614$ and an expected per-seller payoff of approximately $2.74$, compared with the competitive payoff of $2$.

The phase transition at $\rho_c(r)$ (see the vertical line) highlights the critical role of output fidelity. A small increase in output fidelity from just below to just above $\rho_c(r)$ transforms the long-run outcome from purely competitive pricing ($\theta\to 0$) to partial supracompetitive pricing ($\theta\to\theta_+(\rho,r)\ge1/2$). This discontinuity underscores the sensitivity of supracompetitive outcomes to seemingly minor adjustments in model configuration parameters such as sampling temperature or top-$p$ filtering.

Finally, we note that perfect output fidelity maximizes the correlation between sellers' actions, thereby most effectively sustaining supracompetitive pricing. As $\rho$ decreases toward $\rho_c(r)$, the $\Delta(\theta) > 0$ region narrows and eventually vanishes, indicating that stochastic noise in model outputs disrupts coordination and undermines supracompetitive outcomes. \qed
\end{example}

Example~\ref{ex:collusion-region} illustrates several key insights from Proposition~\ref{prop:convergence-binf}.
First, high output fidelity enables supracompetitive pricing. When the model's output fidelity exceeds the critical threshold $\rho_c(r)$, a supracompetitive equilibrium emerges and can be selected by the learning dynamics. This finding is particularly concerning for practical applications. In high-stakes tasks such as pricing, decision-makers typically configure AI pricing models for robustness and reproducibility, favoring low-randomness settings. This contrasts with creative applications where higher stochasticity is preferred. Such operational choices increase the effective output fidelity $\rho$ (through lower generation noise, greater prompt standardization, and higher adherence to recommendations), thereby inadvertently facilitating supracompetitive pricing.
Similarly, greater seller adherence to the model's recommendations is analogous to higher effective output fidelity, further amplifying the risk of supracompetitive pricing.
The threshold $\rho_c(r)$ provides a quantitative benchmark: maintaining $\rho < \rho_c(r)$ through controlled randomness in recommendations would guarantee competitive long-run outcomes, but this runs counter to standard practices that prioritize decision consistency. Our framework quantifies this tradeoff: the threshold $\rho_c(r)$ specifies exactly how much randomness is needed to eliminate the supracompetitive basin of attraction. Any factor that reduces effective fidelity, whether through generation noise, prompt diversity across sellers, or reduced adherence to a single model's output, pushes the market toward competitive outcomes.

Second, supracompetitive pricing emerges even with a moderate initial propensity. In the high-fidelity regime, Proposition~\ref{prop:convergence-binf} establishes that if the model's initial propensity $\theta_0$ exceeds the threshold $\theta_-(\rho,r)$, which lies below $1/2$, the learning dynamics converge to the collusive equilibrium $\theta_+(\rho,r)$. This condition is likely to be satisfied in practice, where decision-makers typically prompt LLMs with real-world business contexts rather than abstract or hypothetical framings. Two factors contribute to this tendency. First, modern LLMs are pretrained on vast corpora of real-world text, including business literature, case studies, and strategic discussions, which extensively document the profitability of coordinated pricing. This pretraining endows the model with a prior belief that high prices are advantageous when competitors behave similarly. Second, reinforcement learning from human feedback (RLHF) and other alignment procedures tune the model to produce responses that humans find helpful and reasonable, potentially reinforcing such priors if human evaluators or users favor collusive recommendations.

Recent experimental evidence corroborates this prediction. For instance, \citet{robinson2025framing} find that when a strategic game is presented as a ``real-world'' business scenario, LLM cooperation (i.e., collusion) rates approach 98\%, whereas reframing the identical rules as an ``imaginary-world'' scenario causes these rates to plummet to 37\%. Together, these observations suggest that off-the-shelf LLMs deployed for real-world pricing decisions may enter the collusive parameter range at initialization.

Third, the interior nature of the collusive equilibrium has critical implications for detection and regulation. Because $\theta_+(\rho,r) < 1$, prices are elevated on average but not perfectly coordinated: occasional low-price recommendations complicate detection methods that rely on identifying perfect price correlation. This pattern resembles tacit collusion in its observable consequences: supra-competitive outcomes emerge without explicit agreement or perfectly synchronized behavior, posing challenges for antitrust enforcement that traditionally requires evidence of coordination.

However, while the \emph{outcome} resembles tacit collusion, the underlying \emph{mechanism} is fundamentally different from the classical economic theory of collusion in repeated games. In that framework, collusion is sustained through intertemporal strategic incentives: firms weigh discounted future profits from maintaining supra-competitive prices against short-term gains from undercutting, with defections deterred by punishment phases such as reversion to competitive play \citep{harrington2018developing}. The mechanism in our framework is fundamentally different. Collusive pricing emerges not from strategic coordination among sellers but from the structural properties of shared decision-making infrastructure. The shared model serves as what antitrust scholars have termed a ``hub'' in a hub-and-spoke arrangement \citep{stucke2023role}: a common intermediary whose latent preferences induce correlated pricing across sellers without direct communication between them. Sellers do not strategically choose to sustain or deviate from collusive prices. Rather, they adopt recommendations from a shared model whose learning dynamics endogenously produce elevated $p_{HH}$. This structural origin distinguishes shared-model-mediated supracompetitive pricing from both explicit collusion, which requires agreement, and the punishment-based tacit collusion identified in the reinforcement learning literature \citep{calvano2020artificial}, thereby posing novel challenges for existing antitrust frameworks.

\subsection{Finite-Batch Regime}
\label{sec:finite-b}

In contrast to the large-batch limit, where estimation noise vanishes, practical model retraining uses finite batches of interaction data. This introduces stochastic fluctuations that persist throughout the learning process and can alter the system's trajectory. The key distinction from the large-batch regime is that equilibrium \emph{selection} becomes random. Recall from Proposition~\ref{prop:convergence-binf} that in the high-fidelity regime ($\rho > \rho_c(r)$), two stable equilibria coexist: competitive pricing ($\theta = 0$) and collusive pricing ($\theta = \theta_+$). With finite batches, identical initial conditions can lead to either outcome depending on the realized training data. We now analyze this finite-batch regime by returning to the stochastic recursion \eqref{eq:logit-recursion}.

Specifically, we define the martingale-difference noise
\[
M_{n+1}\;\triangleq\;\overline{D}_n^{(b)}-\Delta(\theta_n).
\]
Because the batch-average estimator $\overline{D}_n^{(b)}$ is unbiased, we have
\[
\qquad \mathbb{E}[M_{n+1}\mid \theta_n]=0.
\]
With the definition of $M_{n+1}$, \eqref{eq:logit-recursion} becomes
\begin{equation}\label{eq:SA-form-z}
z_{n+1} = z_n + \gamma_n\bigl(\Delta(\theta_n)+M_{n+1}\bigr),\qquad \theta_n=\sigma(z_n).
\end{equation}
For fixed $\rho\in(1/2,1)$, Lemma~\ref{lem:unbiased-D} implies that $\overline D_n^{(b)}$ (and hence $M_{n+1}$) has bounded second moments with $\mathrm{Var}(M_{n+1}\mid\theta_n)=O(1/b)$.

The recursion \eqref{eq:SA-form-z} has a systematic component $\Delta(\theta_n)$, which drives the propensity toward its long-run value, and a noise component $M_{n+1}$, which introduces random fluctuations around this trend. The systematic component is identical to the large-batch case analyzed in the previous section: the same stable equilibria ($\theta=0$ in the low-fidelity regime, and both $\theta=0$ and $\theta_+$ in the high-fidelity regime) govern the long-run behavior. The batch size $b$ affects only the noise variance, leaving this equilibrium structure unchanged. However, when multiple stable equilibria coexist, noise can drive the system across basin boundaries, rendering the ultimate selection among them random. Standard results from stochastic approximation theory can be applied to establish that the propensity $\theta_n$ converges almost surely to one of the stable equilibria, though which one depends on the realized noise path. The following proposition formalizes this convergence.

\begin{proposition}[{\sc Convergence for Finite Batch Size}]\label{prop:convergence-general}
Fix $r\in(1,2)$, $\rho\in(1/2,1)$, batch size $b\ge 1$, a step-size exponent $\alpha\in(1/2,1]$, and a scaling constant $\eta\in(0,1]$. Let $\gamma_n=\eta/(n+1)^\alpha$ and consider the recursion \eqref{eq:logit-recursion} from any $\theta_0\in(0,1)$. Then $\theta_n$ converges almost surely to a (random) equilibrium of \eqref{eq:ODE}. Specifically:
\begin{enumerate}[(i)]
\item If $\rho\in(1/2,\rho_c(r))$, then $\theta_n \xrightarrow{\emph{a.s.}} 0$.
\item If $\rho=\rho_c(r)$, then $\theta_n \xrightarrow{\emph{a.s.}} \theta_\infty$ where $\theta_\infty \in \{0,1/2\}$.
\item If $\rho\in(\rho_c(r),1)$, then $\theta_n \xrightarrow{\emph{a.s.}} \theta_\infty$ where $\theta_\infty \in \{0,\theta_+(\rho,r)\}$.
\end{enumerate}
\end{proposition}

Proposition~\ref{prop:convergence-general} establishes that finite-batch learning converges almost surely. The argument follows the ODE method for stochastic approximation: the recursion tracks the mean-field ODE, and in one dimension, the limit set must reduce to an ODE equilibrium. Standard nonconvergence results further imply that linearly unstable equilibria are selected with probability zero.
In the \emph{low-fidelity regime} ($\rho<\rho_c(r)$), the unique long-run outcome is competitive pricing ($\theta=0$), so convergence is deterministic despite the stochastic dynamics. At the knife-edge \emph{critical regime} ($\rho=\rho_c(r)$), the ODE admits an additional equilibrium at $\theta=1/2$ with one-sided stability, and finite-batch learning can converge to either $\theta=0$ or $\theta=1/2$ depending on the realized noise path. In the \emph{high-fidelity regime} ($\rho>\rho_c(r)$), both $\theta=0$ and $\theta_+(\rho,r)$ are locally stable, and which equilibrium is selected depends on the realized sequence of stochastic shocks $(M_n)_{n\ge 1}$. This path dependence is the defining feature of the finite-batch regime: identical models with identical initial conditions can converge to different long-run outcomes depending on the random training data encountered during training.

\begin{remark}\label{rem:rho1}
When $\rho=1$, the conditional payoff difference is constant: $\Delta(\theta)=2r-2>0$ for all $\theta\in(0,1)$, so the mean-field ODE \eqref{eq:ODE} has $\theta=1$ as its unique attracting equilibrium. Proposition~\ref{prop:convergence-general} is stated for $\rho<1$ because the IPW estimator \eqref{eq:D-def} becomes unbounded as $\theta\to 0$ or $\theta\to 1$ when $\rho=1$ (i.e., an ``off-policy without exploration'' issue). This can be addressed by mild clipping or enforced exploration, replacing $p_H(\theta)$ and $p_L(\theta)$ in \eqref{eq:D-def} with $\max\{p_H(\theta),\varepsilon\}$ and $\max\{p_L(\theta),\varepsilon\}$ for a small $\varepsilon>0$. Under such standard stabilizations, the stochastic recursion continues to track the same ODE and converges to $\theta=1$, for $\rho=1$ as well.
\end{remark}

The preceding analysis shows that convergence occurs but leaves open the question of \emph{which} equilibrium is selected in the high-fidelity regime. The following proposition characterizes this equilibrium selection and clarifies how the batch size governs the probability of selection.

\begin{proposition}[{\sc Equilibrium Selection and the Effect of Batch Size}]\label{prop:batch-selection}
Fix $r\in(1,2)$ and $\rho\in(\rho_c(r),1)$, and let $\theta_-(\rho,r)<\theta_+(\rho,r)$ denote the unstable and stable interior equilibria from Proposition~\ref{prop:H-dominance}.
Fix step sizes $\gamma_n=\eta/(n+1)^\alpha$ with $\alpha\in(1/2,1]$ and $\eta\in(0,1]$, and consider the stochastic recursion \eqref{eq:logit-recursion} with batch size $b\ge 1$ from an initial condition $\theta_0\in(0,1)$.
Define the collusion probability
\[
p_+(b,\theta_0)\;\triangleq\;\mathbb{P}\Bigl(\lim_{n\to\infty}\theta_n=\theta_+(\rho,r)\Bigr).
\]

Then the following holds.

\begin{enumerate}[(i)]
\item {\sc (Finite-Horizon Tracking with Explicit $b$ Dependence).}
For every $T>0$ and every $\varepsilon>0$, there exists a constant $C=C(T,\rho,r,\alpha,\eta)<\infty$, independent of $b$, such that
\[
\mathbb{P}\Bigl(\sup_{0\le t\le T}\bigl|\bar\theta(t)-\theta^{\det}(t)\bigr|>\varepsilon\Bigr)\;\le\;\frac{C}{b\,\varepsilon^2},
\]
where $\bar\theta(t)\triangleq\sigma(\bar z(t))$ is the continuous-time interpolation of the stochastic sequence $(z_n)$ (as in Lemma~\ref{lem:ODE-tracking})
and $\theta^{\det}(t)$ is the continuous-time interpolation of the deterministic (large-batch) recursion \eqref{eq:deterministic-update-z} started from the same $\theta_0$.
In particular, $\sup_{0\le t\le T}|\bar\theta(t)-\theta^{\det}(t)|=O_p(1/\sqrt b)$.

\item {\sc (Lock-In away from the Separatrix).}
Fix $\delta\in(0,1/2)$ and define
\[
K_+(\delta)\triangleq[\theta_-+\delta,\,1-\delta],
\qquad
K_-(\delta)\triangleq[\delta,\,\theta_- - \delta].
\]
There exist constants $c_\delta>0$ and $C_\delta<\infty$ such that, for all $b\ge 1$,
\[
\sup_{\theta_0\in K_+(\delta)}\bigl(1-p_+(b,\theta_0)\bigr)\;\le\; C_\delta\,e^{-c_\delta\,b\,\delta^2},
\qquad
\sup_{\theta_0\in K_-(\delta)}p_+(b,\theta_0)\;\le\; C_\delta\,e^{-c_\delta\,b\,\delta^2}.
\]
Consequently, for any fixed $\theta_0\neq\theta_-$ we have
\[
\lim_{b\to\infty}p_+(b,\theta_0)=\mathbbm{1}\{\theta_0>\theta_-\}.
\]

\item {\sc (Transition Region Width).}
For every $\epsilon\in(0,1/2)$ and every $\delta_0\in(0,1/2)$ with $\theta_-\in(\delta_0,1-\delta_0)$, there exists $K=K(\rho,r,\alpha,\eta,\epsilon,\delta_0)>0$ such that for all $b\ge 1$ and all $\theta_0\in[\delta_0,1-\delta_0]$,
\[
p_+(b,\theta_0)\in(\epsilon,1-\epsilon)\quad\Longrightarrow\quad
|\theta_0-\theta_-|\;\le\;\frac{K}{\sqrt b}.
\]
\end{enumerate}
\end{proposition}

Proposition~\ref{prop:batch-selection} makes precise that equilibrium selection in the finite-batch regime is governed by two forces: deterministic drift and batch-induced noise.
Part~(i) quantifies finite-horizon tracking of the large-batch recursion, with deviations scaling as $O_p(1/\sqrt b)$.
Part~(ii) turns this tracking into an equilibrium selection statement: starting a fixed margin $\delta$ above (below) the unstable threshold $\theta_-$, the probability of selecting the ``wrong'' equilibrium decays at rate $e^{-c b\delta^2}$.
The proof of Part~(ii) is a one-dimensional barrier argument: reaching the wrong basin requires the weighted noise martingale to overcome a deterministic log-odds gap of order $\delta$.
Part~(iii) summarizes the implication for comparative statics: the only initial conditions that yield a nontrivial mixture of competitive and collusive outcomes lie in an $O(1/\sqrt b)$ neighborhood of $\theta_-$.


\begin{example}[{\sc Batch Size and Collusion Probability}]\label{ex:batch-size}
We illustrate the effect of the batch size on equilibrium selection using the same parameter setting as Example~\ref{ex:collusion-region}: $r=1.5$ and $\theta_0=1/2$. We set $\rho=0.85$, which exceeds the critical threshold $\rho_c(1.5)\approx 0.789$. For these parameters, $\theta_-\approx 0.154$ and $\theta_+\approx 0.846$. Since the initial propensity $\theta_0=1/2$ exceeds $\theta_-$, the deterministic dynamics (i.e., the large-batch limit) would converge to the collusive equilibrium $\theta_+$. Following standard practice in the stochastic approximation literature, we set the step size to $\gamma_n = 1/(n+1)^\alpha$. We note that condition (A1) requires $\alpha\in(1/2,1]$. We thus set $\alpha=2/3$ to ensure practical convergence within the simulation horizon.
Figure~\ref{fig:batch-size-effect} displays 20 independent trajectories of $\theta_n$ for batch sizes $b\in\{1,4,16,64\}$, along with their sample means. 

Several patterns emerge. First, all trajectories across all batch sizes eventually stabilize, consistent with the almost-sure convergence established in Proposition~\ref{prop:convergence-general}. Second, the variance of the trajectories decreases markedly with the batch size. For $b=1$, trajectories exhibit substantial volatility and frequently cross the unstable threshold $\theta_-$. For $b=64$, trajectories remain tightly concentrated around the mean path and converge to the theoretical equilibrium $\theta_+\approx 0.846$. Third, the fraction of trajectories converging to the collusive equilibrium $\theta_+$ increases with the batch size in this simulation: 15\% for $b=1$, 70\% for $b=4$, 95\% for $b=16$, and 100\% for $b=64$. This pattern is consistent with Proposition~\ref{prop:batch-selection}, which predicts that when $\theta_0>\theta_-$, the collusion probability rises toward one as batch noise vanishes. \qed

\begin{figure}[!htb]
    \centering
    \includegraphics[width=0.8\textwidth]{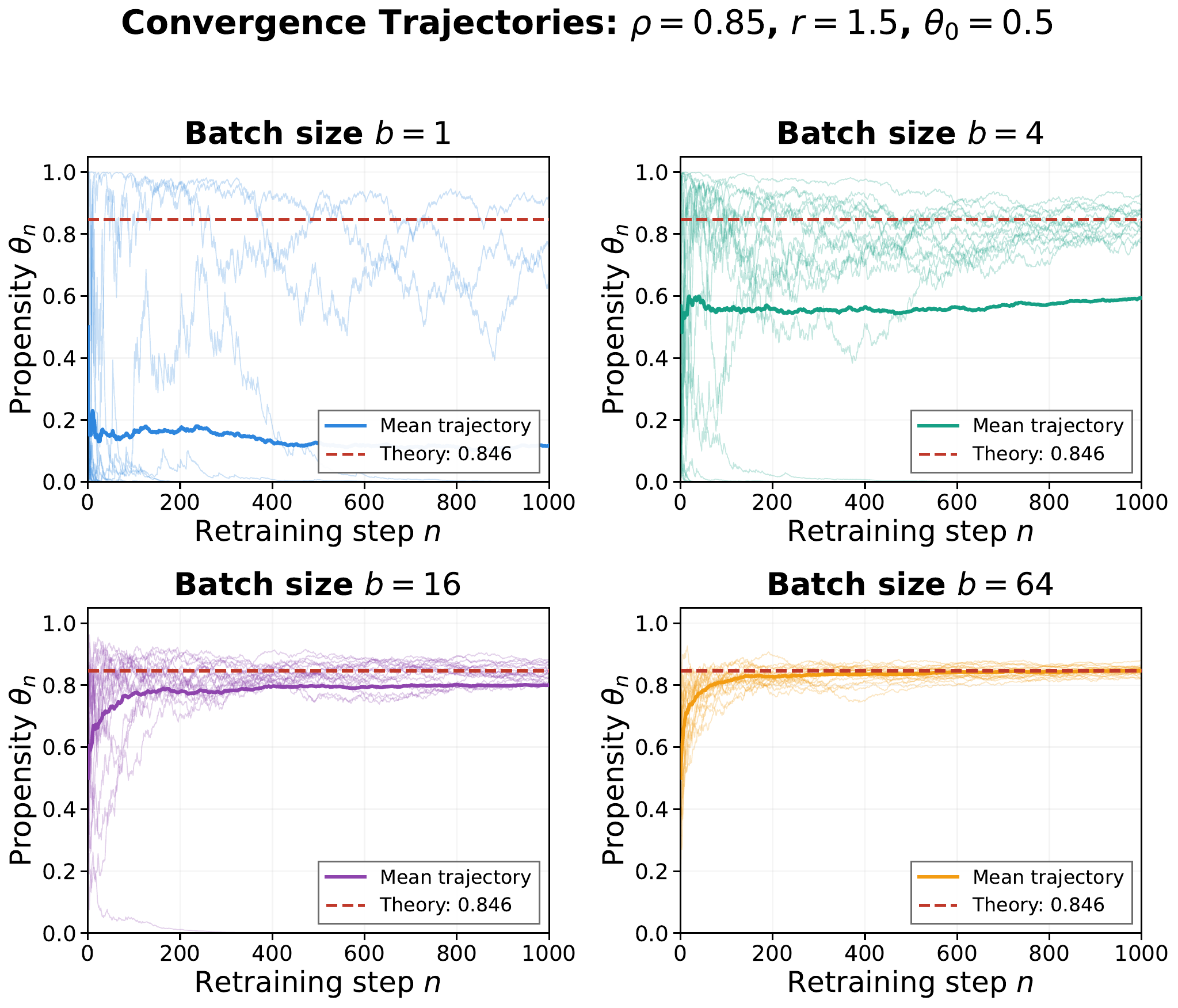}
    \caption{Convergence trajectories of propensity $\theta_n$ for different batch sizes. Each panel displays 20 independent replications (faint lines) and their sample mean (bold line) for the batch size $b\in\{1,4,16,64\}$. The horizontal dashed line indicates the theoretical collusive equilibrium $\theta_+(\rho,r)\approx 0.846$. Parameters: $\rho=0.85$, $r=1.5$, $\theta_0=1/2$. Larger batch sizes reduce trajectory variance and increase the probability of convergence to the collusive outcome.}
    \label{fig:batch-size-effect}
\end{figure}
\end{example}

Example~\ref{ex:batch-size} carries important implications for understanding supracompetitive pricing risk in practice. Proposition~\ref{prop:batch-selection} implies that when the initial propensity lies above the unstable threshold $\theta_-$, increasing the batch size suppresses noise and pushes the selection probability closer to the large-batch prediction of collusion. In the context of LLM-based pricing, the batch size $b$ corresponds to the number of pricing rounds accumulated before each retraining update.
In practice, although major LLM providers have accelerated their release cadence (for example, OpenAI released GPT-4.5 in February 2025 and GPT-5 in August 2025, Anthropic released Claude 4 in May 2025 and Claude 4.5 in September 2025, and Google released Gemini 2.5 in March 2025 and Gemini 3.0 in November 2025), effective batch sizes remain large because each retraining cycle aggregates pricing interactions from many sellers across many rounds. The widespread adoption of shared AI models for pricing means that even a single update cycle accumulates a substantial volume of outcome data. Between major releases, models may receive incremental updates via parameter-efficient fine-tuning or retrieval augmentation, but the core model weights, which encode the propensity $\theta$ in our model, remain fixed mainly.
This deployment-scale reality translates directly into large effective batch sizes.

Coupled with the fact that real-world LLMs likely possess initial propensities $\theta_0>\theta_-$ owing to pretraining on business corpora and RLHF alignment, Proposition~\ref{prop:batch-selection} delivers a striking implication: larger effective batch sizes can increase supracompetitive pricing risk by suppressing the noise that might otherwise push the model across the unstable threshold toward competitive pricing. In practice, even as providers increase their update frequency, effective batch sizes remain large because each update cycle aggregates interactions from many sellers across many pricing rounds.

\section{Conclusion}\label{sec:conclusion}

This paper develops a theoretical framework to analyze the emergence of supracompetitive pricing when competing sellers delegate pricing decisions to a shared AI model. Our analysis identifies a fundamental mechanism: the model's common pricing propensity creates correlated recommendations across sellers, and a learning process that updates this propensity based on observed performance can drive prices above competitive levels without any explicit coordination.

Our framework is subject to several limitations that suggest directions for future research. First, the model considers a duopoly setting with two sellers. Extending the analysis to $N$ sellers introduces two opposing forces: more sellers increase the probability of at least one recommendation miscoordination, and miscoordinations favor low-price sellers, potentially undermining collusion. Simultaneously, however, the shared model creates market-wide correlation, as all $N$ sellers receive signals drawn from the same latent preference, amplifying the monoculture effect. A central research question can be how the critical fidelity threshold $\rho_c$ varies with $N$: Does market fragmentation protect against or amplify AI-induced supracompetitive pricing?

Second, our model assumes that sellers query the model without historical context, so both receive recommendations drawn from the same propensity $\theta$. In practice, queries may incorporate historical pricing data, sales outcomes, or competitor behavior. With such context, the model could adapt its recommendations to each seller's history through in-context learning \citep{brown2020language}, potentially providing different policies to different sellers within each batch. One seller might receive signals that push toward collusive pricing, while another receives signals that push toward competitive pricing. When these divergent experiences are merged during retraining, the key question is whether policies converge to a common equilibrium or persistently remain heterogeneous. This connects to cutting-edge topics in machine learning, including continual learning, in which models must learn from sequential, potentially conflicting data streams \citep{parisi2019continual}, and catastrophic forgetting, in which training on new data overwrites previously learned knowledge \citep{kirkpatrick2017overcoming}. Understanding how these dynamics interact with the collusion mechanisms identified in this paper presents an important direction for future research.


\def\bibsep{-2.5pt}
\bibliographystyle{ormsv080}
\bibliography{main}

\newpage
\setcounter{page}{1}
\ECSwitch
\begin{center}
 {\bf\large
 Online Appendix to \\ ``Supracompetitive Pricing Under AI Monoculture'':\\
 Proofs
  }
\end{center}
\renewcommand\thesection{\Alph{section}}
\renewcommand\theHsection{\Alph{section}}

\setcounter{equation}{0}
\setcounter{section}{0}
\setcounter{proposition}{0}
\setcounter{figure}{0}
\setcounter{table}{0}

\numberwithin{equation}{section}
\numberwithin{proposition}{section}
\numberwithin{figure}{section}
\numberwithin{table}{section}


\noindent \textbf{Proof of Lemma \ref{lem:unbiased-D}.}
For fixed $\theta$, by the law of total expectation,
\begin{align*}
\mathbb{E}[D(\theta)\mid\theta]
&= \frac{1}{p_H(\theta)}\mathbb{E}[\Pi\,\mathbbm{1}\{S=H\}] - \frac{1}{p_L(\theta)}\mathbb{E}[\Pi\,\mathbbm{1}\{S=L\}]\\
&= \frac{\mathbb{P}(S=H)}{p_H(\theta)}\mathbb{E}[\Pi \mid S=H] - \frac{\mathbb{P}(S=L)}{p_L(\theta)}\mathbb{E}[\Pi \mid S=L]\\
&= \mathbb{E}[\Pi \mid S=H] - \mathbb{E}[\Pi \mid S=L] \;=\; \Delta(\theta).
\end{align*}
For the boundedness claim, note that payoffs are bounded and, for any fixed $\rho\in(1/2,1)$,
\[
\inf_{\theta\in[0,1]}\min\{p_H(\theta),p_L(\theta)\} = 1-\rho \;>\;0,
\]
and with $\Pi_{\max}\triangleq \max\{r, 2r,2+r,2\}=2+r$, we have $\bigl|D(\theta)\bigr|\le \Pi_{\max}/(1-\rho)<\infty$ uniformly in $\theta$.
\qed

\vskip 0.5cm
\noindent \textbf{Proof of Lemma \ref{lem:ODE-tracking}.}
We organize the proof into six steps. Steps 1 to 5 establish the tracking statement in part (i) for the log-odds recursion. Step 6 works in propensity space to conclude part (ii).

\paragraph{Step 1: Drift-noise decomposition.}
Rewrite the recursion \eqref{eq:logit-recursion} as
\begin{equation}\label{eq:SA-decomposition}
z_{n+1} = z_n + \gamma_n \bigl[\Delta(\theta_n) + M_{n+1}\bigr],
\end{equation}
where $\theta_n = \sigma(z_n)$ and the noise term is defined as
\[
M_{n+1} := \overline{D}_n^{(b)} - \Delta(\theta_n).
\]
By assumption (A2), the sequence $(M_{n+1})_{n \geq 0}$ is a martingale difference sequence adapted to the filtration $(\mathcal{H}_n)_{n \geq 0}$:
\[
\mathbb{E}[M_{n+1} \mid \mathcal{H}_n] = \mathbb{E}[\overline{D}_n^{(b)} \mid \mathcal{H}_n] - \Delta(\theta_n) = 0.
\]
Moreover, by assumption (A3) and the boundedness of $\Delta(\cdot)$ established in Lemma~\ref{lem:unbiased-D}, the second moments are uniformly bounded:
\begin{equation}\label{eq:noise-bound}
\sup_{n \geq 0} \mathbb{E}[M_{n+1}^2] \le 2\sup_{n}\mathbb{E}[(\overline{D}_n^{(b)})^2] + 2\sup_{\theta \in [0,1]}|\Delta(\theta)|^2 < \infty.
\end{equation}

\paragraph{Step 2: Almost-sure convergence of the noise sum.}
Define the partial sums $W_n := \sum_{k=0}^{n-1} \gamma_k M_{k+1}$. We claim that $(W_n)_{n \geq 0}$ converges almost surely to a finite limit.

To see this, observe that $(W_n,\mathcal{H}_n)_{n \geq 0}$ is a martingale and
\[
\mathbb{E}[W_n^2] = \sum_{k=0}^{n-1} \gamma_k^2 \, \mathbb{E}[M_{k+1}^2] \le C_M \sum_{k=0}^{n-1} \gamma_k^2,
\]
where $C_M := \sup_k \mathbb{E}[M_{k+1}^2] < \infty$ by \eqref{eq:noise-bound}. Since $\sum_{k=0}^\infty \gamma_k^2 < \infty$, the martingale $(W_n)$ is $L^2$-bounded. By the martingale convergence theorem \citep[Theorem~5.4.9]{durrett2019}, $W_n$ converges almost surely (and in $L^2$) to a finite limit $W_\infty$.

\paragraph{Step 3: Regularity of the drift function.}
Define $h: \mathbb{R} \to \mathbb{R}$ by $h(z) := \Delta(\sigma(z))$. We verify that $h$ is globally Lipschitz and bounded.

The sigmoid function $\sigma(z) = (1 + e^{-z})^{-1}$ is infinitely differentiable with
\[
\sigma'(z) = \sigma(z)\bigl(1 - \sigma(z)\bigr) \in (0, 1/4] \quad \text{for all } z \in \mathbb{R}.
\]
By construction, $\Delta(\theta)$ admits an explicit rational expression in $\theta$, with denominator $p_H(\theta)p_L(\theta)$. For $\rho \in (1/2, 1)$, Lemma~\ref{lem:unbiased-D} shows that $p_H(\theta), p_L(\theta) \geq 1 - \rho > 0$ for all $\theta \in [0,1]$, so $\Delta$ is continuously differentiable on $[0,1]$ and $\sup_{\theta \in [0,1]}|\Delta'(\theta)| < \infty$. Consequently, $h = \Delta \circ \sigma$ is continuously differentiable on $\mathbb{R}$ and
\[
|h'(z)| = \bigl|\Delta'(\sigma(z))\bigr|\,\sigma'(z) \le \frac{1}{4}\sup_{\theta \in [0,1]}|\Delta'(\theta)|,
\]
which implies a global Lipschitz constant. In addition, $|h(z)| \le \sup_{\theta \in [0,1]}|\Delta(\theta)|$ for all $z$.

\paragraph{Step 4: Discrete-to-continuous comparison via Gronwall's inequality.}
Fix a horizon $T > 0$. For any initial condition $z \in \mathbb{R}$, let $\Phi_t(z)$ denote the solution to the ODE $\dot{z} = h(z)$ at time $t$ when starting from $z(0) = z$. We compare the discrete sequence $(z_n)$ to the ODE trajectory $(\Phi_{t-t_n}(z_n))_{t \geq t_n}$.

For $t \in [t_n, t_{n+1}]$, the ODE solution starting from $z_n$ satisfies
\[
\Phi_{t-t_n}(z_n) = z_n + \int_0^{t-t_n} h\bigl(\Phi_s(z_n)\bigr) \, ds.
\]
In particular, setting $H := \sup_z |h(z)|$ and $L$ to be a global Lipschitz constant for $h$, we have
\[
\bigl|\Phi_{\gamma_n}(z_n) - z_n - \gamma_n h(z_n)\bigr|
= \biggl|\int_0^{\gamma_n}\!\!\bigl(h(\Phi_s(z_n)) - h(z_n)\bigr)\,ds\biggr|
\le \int_0^{\gamma_n} L \,|\Phi_s(z_n) - z_n|\,ds
\le \int_0^{\gamma_n} L s H\,ds
= \frac{LH}{2}\,\gamma_n^2.
\]

Combining this with \eqref{eq:SA-decomposition}, the one-step error $\varepsilon_n := z_{n+1} - \Phi_{\gamma_n}(z_n)$ satisfies
\[
|\varepsilon_n| \le \gamma_n |M_{n+1}| + \frac{LH}{2}\,\gamma_n^2.
\]
Iterating and applying a discrete Gronwall argument yields a uniform bound for the deviation between the interpolated path and the ODE trajectory over $[t_n,t_n+T]$ in terms of the martingale increments $W_m-W_n$ and the tail sum $\sum_{k=n}^\infty \gamma_k^2$.

\paragraph{Step 5: Tracking error decay.}
Define the tracking error over a fixed time horizon $T > 0$ as
\[
V_n \;:=\; \sup_{0 \leq t \leq T} \bigl|\bar{z}(t_n + t) - \Phi_t(z_n)\bigr|.
\]
Let $N_T(n)$ denote the number of discrete steps in the time window $[t_n, t_n + T]$. Since $h$ is globally Lipschitz and bounded, applying the discrete Gronwall inequality to the one-step errors from Step~4 yields a constant $C_T < \infty$ (depending only on $T$, $L$, and $H$) such that
\[
V_n \le C_T \biggl( \sup_{m \ge n}\bigl|W_m - W_n\bigr| + \sum_{k=n}^{\infty} \gamma_k^2 \biggr).
\]
From Step~2, the partial sums $W_m := \sum_{k=0}^{m-1} \gamma_k M_{k+1}$ converge almost surely to a finite limit $W_\infty$. Therefore,
\[
\sup_{m \ge n}\bigl|W_m - W_n\bigr| \to 0 \quad \text{a.s.}
\]
as $n \to \infty$. Since $\sum_{k=0}^\infty \gamma_k^2 < \infty$, the tail sum also vanishes: $\sum_{k=n}^\infty \gamma_k^2 \to 0$. Combining these estimates yields $V_n \to 0$ almost surely, establishing part (i).

\paragraph{Step 6: Limit set characterization.}
Let $z^\star$ be a limit point of $(z_n)$ and let $z_{n_j} \to z^\star$ along a subsequence. Then $\theta_{n_j} := \sigma(z_{n_j}) \to \theta^\star := \sigma(z^\star)$ with $\theta^\star \in (0,1)$. We now work in propensity space.

By a second-order Taylor expansion of $\sigma$, there exists $\xi_n$ between $z_n$ and $z_{n+1}$ such that
\[
\theta_{n+1}
= \theta_n + \gamma_n \sigma'(z_n)\,\overline{D}_n^{(b)} + \frac{1}{2}\sigma''(\xi_n)\,\gamma_n^2\bigl(\overline{D}_n^{(b)}\bigr)^2
= \theta_n + \gamma_n \theta_n(1-\theta_n)\,\overline{D}_n^{(b)} + \frac{1}{2}\sigma''(\xi_n)\,\gamma_n^2\bigl(\overline{D}_n^{(b)}\bigr)^2.
\]
Write $\overline{D}_n^{(b)} = \Delta(\theta_n) + M_{n+1}$ and define
\[
F(\theta) := \theta(1-\theta)\Delta(\theta),\qquad \widetilde M_{n+1} := \theta_n(1-\theta_n)M_{n+1},\qquad r_{n+1} := \frac{1}{2}\sigma''(\xi_n)\,\gamma_n \bigl(\overline{D}_n^{(b)}\bigr)^2.
\]
Then the propensity recursion becomes
\[
\theta_{n+1} = \theta_n + \gamma_n\bigl(F(\theta_n) + \widetilde M_{n+1} + r_{n+1}\bigr).
\]
Since $|\sigma''(z)| = |\sigma(z)(1-\sigma(z))(1-2\sigma(z))| \le 1/4$ for all $z$, we have
\[
\sum_{n=0}^\infty \gamma_n |r_{n+1}|
\le \frac{1}{8}\sum_{n=0}^\infty \gamma_n^2 \bigl(\overline{D}_n^{(b)}\bigr)^2
< \infty \quad \text{a.s.},
\]
where the almost-sure finiteness follows from (A1) and (A3) by Fubini's theorem. Moreover, $(\widetilde M_{n+1})$ is a martingale difference sequence with uniformly bounded second moments since $\theta_n(1-\theta_n) \le 1/4$.

Standard stochastic approximation results imply that the piecewise-linear interpolation of $(\theta_n)$ is an asymptotic pseudotrajectory of the ODE $\dot{\theta} = F(\theta)$, which is \eqref{eq:ODE}. Since $(\theta_n)$ is bounded, \citet[Theorem~5.7]{benaim1999} implies that its $\omega$-limit set is internally chain transitive for this one-dimensional flow. In one dimension, any internally chain transitive set must be contained in the equilibrium set $\{\theta : F(\theta)=0\}$. Therefore, every limit point $\theta^\star$ satisfies $F(\theta^\star)=0$. For $\theta^\star \in (0,1)$ this reduces to $\Delta(\theta^\star)=0$, which is equivalent to $\Delta(\sigma(z^\star))=0$. This establishes part (ii) and completes the proof.
\qed

\vskip 0.5cm
\noindent \textbf{Proof of Proposition \ref{prop:H-dominance}.}
Let $S_i\in\{H,L\}$ denote the recommendation to Seller~$i$ and let $\Pi_i$ denote Seller~$i$'s profit in a single period. From Table~\ref{tab:payoff_matrix}:
\begin{align*}
&(H,H): &&\Pi_1 = 2r,\quad \Pi_2 = 2r,\\
&(H,L): &&\Pi_1 = r,\quad \Pi_2 = 2+r,\\
&(L,H): &&\Pi_1 = 2+r,\quad \Pi_2 = r,\\
&(L,L): &&\Pi_1 = 2,\quad \Pi_2 = 2.
\end{align*}

Using the joint probabilities \eqref{eq:pHH}--\eqref{eq:pLL}, the conditional expected profits are
\begin{align}
\mathbb{E}[\Pi_1 \mid S_1 = H, \theta] &= \frac{p_{HH}(\theta) \cdot 2r + p_{HL}(\theta) \cdot r}{p_{HH}(\theta) + p_{HL}(\theta)}, \label{eq:EH-cond-app}\\
\mathbb{E}[\Pi_1 \mid S_1 = L, \theta] &= \frac{p_{LH}(\theta) \cdot (2+r) + p_{LL}(\theta) \cdot 2}{p_{LH}(\theta) + p_{LL}(\theta)}. \label{eq:EL-cond-app}
\end{align}
The same expressions hold for Seller~2 by symmetry.
Let $A:=p_{HH}(\theta)$, $B:=p_{LL}(\theta)$, and $q:=p_{HL}(\theta)=p_{LH}(\theta)=\rho(1-\rho)$. The condition $\Delta(\theta)\ge 0$ is equivalent to
\begin{equation}\label{eq:ineq-cross-app}
r(2A+q)(q+B) \;\ge\; (A+q)\bigl((r+2)q + 2B\bigr),
\end{equation}
which follows by cross-multiplying \eqref{eq:EH-cond-app}--\eqref{eq:EL-cond-app}.

Let $\kappa\triangleq \rho^2-(1-\rho)^2 = 2\rho-1$. Then
\[
A=(1-\rho)^2+\kappa\theta,\qquad B=\rho^2-\kappa\theta.
\]
Substituting these expressions into \eqref{eq:ineq-cross-app} and simplifying yields
\begin{equation}\label{eq:theta-quadratic-app}
\theta(1-\theta) \;\ge\; K(\rho,r) \;\triangleq\; \frac{(2-r)\,\rho(1-\rho)}{2(r-1)(2\rho-1)^2}.
\end{equation}

Define $s(\rho,r)\triangleq 4K(\rho,r)$, as in \eqref{eq:s-definition}. Then \eqref{eq:theta-quadratic-app} is equivalent to $\theta(1-\theta)\ge s(\rho,r)/4$.

The function $f(\theta)=\theta(1-\theta)$ is concave on $[0,1]$ and attains its maximum $1/4$ at $\theta=1/2$. Consider
\[
g(\theta)\triangleq \theta^2-\theta+\frac{s(\rho,r)}{4}.
\]
Then $\theta(1-\theta)\ge s/4$ is equivalent to $g(\theta)\le 0$. The discriminant of $g$ is $1-s(\rho,r)$.

\paragraph{Case $s(\rho,r)>1$.} Then $1-s<0$ and $g(\theta)>0$ for all $\theta$, so \eqref{eq:theta-quadratic-app} never holds and therefore $\Delta(\theta)<0$ for all $\theta\in(0,1)$.

\paragraph{Case $s(\rho,r)=1$.} Then $g(\theta)=(\theta-1/2)^2$, so \eqref{eq:theta-quadratic-app} holds if and only if $\theta=1/2$, implying $\Delta(1/2)=0$ and $\Delta(\theta)<0$ for $\theta\ne 1/2$.

\paragraph{Case $s(\rho,r)<1$.} Then $g$ has two real roots
\[
\theta_\pm=\frac{1}{2}\pm \frac{1}{2}\sqrt{1-s(\rho,r)},
\]
and $g(\theta)\le 0$ if and only if $\theta\in[\theta_-,\theta_+]$. This implies $\Delta(\theta)>0$ for $\theta\in(\theta_-,\theta_+)$, with $\Delta(\theta_\pm)=0$ and $\Delta(\theta)<0$ outside $[\theta_-,\theta_+]$.
\qed

\vskip 0.5cm
\noindent \textbf{Proof of Proposition \ref{prop:convergence-binf}.}
The proof proceeds in three parts: (i) establishing almost-sure convergence of the stochastic recursion to the limiting deterministic recursion as $b \to \infty$; (ii) deriving the critical fidelity threshold $\rho_c(r)$; and (iii) characterizing the long-run behavior of the limiting recursion for all initial conditions $\theta_0 \in (0,1)$.

\paragraph{Part I: Convergence to the limiting recursion as $b \to \infty$.}

Let $(\theta_n^{(b)})_{n \geq 0}$ denote the sequence generated by the stochastic recursion \eqref{eq:logit-recursion} with batch size $b$, and let $(\theta_n)_{n \geq 0}$ denote the solution of the limiting recursion \eqref{eq:deterministic-update-z} with the same initial condition $\theta_0$.

For each $n \geq 0$, define the estimation error $\varepsilon_n^{(b)} := \overline{D}_n^{(b)} - \Delta(\theta_n^{(b)})$. By Lemma~\ref{lem:unbiased-D}, conditional on the history $\mathcal{H}_n$, the estimator $\overline{D}_n^{(b)}$ is an unbiased estimator of $\Delta(\theta_n^{(b)})$ with variance $\mathrm{Var}(\overline{D}_n^{(b)} \mid \mathcal{H}_n) = O(1/b)$. More precisely, for any fixed $\rho \in (1/2, 1)$, there exists a constant $C_\rho < \infty$ such that $\mathrm{Var}(\overline{D}_n^{(b)} \mid \mathcal{H}_n) \leq C_\rho / b$ uniformly in $n$ and $\theta_n^{(b)} \in (0,1)$.

We now show that $\theta_n^{(b)} \to \theta_n$ almost surely as $b \to \infty$, uniformly over any finite horizon. Fix $N < \infty$. The stochastic recursion can be written as
\[
z_{n+1}^{(b)} = z_n^{(b)} + \gamma_n \bigl[\Delta(\theta_n^{(b)}) + \varepsilon_n^{(b)}\bigr],
\]
while the limiting recursion satisfies
\[
z_{n+1} = z_n + \gamma_n \Delta(\theta_n).
\]
Taking differences and using the global Lipschitz continuity of $h(z) := \Delta(\sigma(z))$ from Lemma~\ref{lem:ODE-tracking}, we obtain
\[
|z_{n+1}^{(b)} - z_{n+1}| \leq |z_n^{(b)} - z_n| + \gamma_n L |z_n^{(b)} - z_n| + \gamma_n |\varepsilon_n^{(b)}|,
\]
where $L$ is a global Lipschitz constant of $h$.

Iterating this bound and applying the discrete Gronwall inequality yields
\[
\max_{0 \leq n \leq N} |z_n^{(b)} - z_n| \leq \exp\Bigl(L \sum_{k=0}^{N-1} \gamma_k\Bigr) \sum_{k=0}^{N-1} \gamma_k |\varepsilon_k^{(b)}|.
\]
To upgrade to almost sure convergence in $b$, we work under a natural coupling where, at each step $k$, the batch of size $b$ is the prefix of an infinite i.i.d. stream of interaction rounds. We then use the uniform boundedness of the per-round estimator. For fixed $\rho\in(1/2,1)$, Lemma~\ref{lem:unbiased-D} implies that $D(\theta)$ is uniformly bounded over $\theta\in[0,1]$, so there exists $B_\rho<\infty$ such that $|D(\theta)|\le B_\rho$ almost surely for all $\theta\in[0,1]$. Consequently, conditional on $\mathcal H_k$, the estimation error $\varepsilon_k^{(b)}$ is an average of $b$ independent, mean-zero, bounded random variables, and Hoeffding's inequality yields a constant $c_\rho>0$ such that for any $\delta>0$,
\[
\mathbb{P}\Bigl(|\varepsilon_k^{(b)}|>\delta \,\big|\, \mathcal H_k\Bigr)\le 2\exp\bigl(-c_\rho b\delta^2\bigr).
\]
Taking expectations and applying a union bound over $k=0,1,\ldots,N-1$ gives
\[
\mathbb{P}\Bigl(\max_{0\le k\le N-1}|\varepsilon_k^{(b)}|>b^{-1/3}\Bigr)
\le 2N\exp\bigl(-c_\rho b^{1/3}\bigr).
\]
Since $\sum_{b=1}^\infty 2N\exp(-c_\rho b^{1/3})<\infty$, the Borel--Cantelli lemma implies that $\max_{0\le k\le N-1}|\varepsilon_k^{(b)}|\le b^{-1/3}$ eventually almost surely as $b\to\infty$. On this event,
\[
\max_{0 \leq n \leq N} |z_n^{(b)} - z_n|
\le \exp\Bigl(L \sum_{k=0}^{N-1} \gamma_k\Bigr)\,b^{-1/3}\sum_{k=0}^{N-1}\gamma_k
\xrightarrow{b\to\infty} 0
\]
almost surely. Since $\sigma$ is Lipschitz with constant $1/4$, this implies $\max_{0 \leq n \leq N} |\theta_n^{(b)} - \theta_n|\to 0$ almost surely. Since $N$ was arbitrary, a diagonalization argument yields almost-sure convergence uniformly on finite horizons.

\paragraph{Part II: Derivation of the critical fidelity threshold.}

By Proposition~\ref{prop:H-dominance}, the sign of $\Delta(\theta)$ on the interior $(0,1)$ is determined by the statistic $s(\rho, r)$ defined in \eqref{eq:s-definition}. The critical case, where $\Delta(\theta) = 0$ has a unique interior solution at $\theta = 1/2$, occurs when $s(\rho, r) = 1$, i.e.,
\[
\frac{2(2-r)\rho(1-\rho)}{(r-1)(2\rho-1)^2} = 1.
\]
Cross-multiplying and rearranging yields the quadratic equation
\[
2(2-r)\rho(1-\rho) = (r-1)(2\rho - 1)^2.
\]
Expanding both sides:
\begin{align*}
\text{LHS} &= 2(2-r)\rho - 2(2-r)\rho^2 = (4 - 2r)\rho - (4 - 2r)\rho^2, \\
\text{RHS} &= (r-1)(4\rho^2 - 4\rho + 1) = (4r - 4)\rho^2 - (4r - 4)\rho + (r - 1).
\end{align*}
Collecting terms:
\[
0 = \bigl[(4r - 4) + (4 - 2r)\bigr]\rho^2 - \bigl[(4r - 4) + (4 - 2r)\bigr]\rho + (r - 1).
\]
Simplifying the coefficient: $(4r - 4) + (4 - 2r) = 2r$, so
\[
2r\rho^2 - 2r\rho + (r - 1) = 0.
\]
Dividing by $2r$ and applying the quadratic formula:
\[
\rho = \frac{1 \pm \sqrt{1 - 2(r-1)/r}}{2} = \frac{1 \pm \sqrt{(2-r)/r}}{2}.
\]
Since $\rho \in (1/2, 1]$ and $r \in (1, 2)$ implies $(2-r)/r \in (0, 1)$, we have $\sqrt{(2-r)/r} \in (0, 1)$. The solution $\rho_- = (1 - \sqrt{(2-r)/r})/2 < 1/2$ lies outside the feasible range, so the unique critical threshold is
\[
\rho_c(r) = \frac{1 + \sqrt{(2-r)/r}}{2}.
\]

\paragraph{Part III: Long-run behavior of the limiting recursion.}

We now characterize the convergence of $(\theta_n)_{n \geq 0}$ as $n \to \infty$ for each regime of $\rho$.

Write $h(z)\triangleq \Delta(\sigma(z))$, so the limiting recursion is $z_{n+1}=z_n+\gamma_n h(z_n)$ with $\theta_n=\sigma(z_n)$. Under our step-size choice, $0<\gamma_n\le \eta\le 1$.

We first record a monotonicity property that prevents spurious basin crossings in the deterministic recursion. For $\rho\in(1/2,1)$, one can simplify the payoff difference to
\[
\Delta(\theta)=2(r-1)-\frac{r\rho(1-\rho)}{p_H(\theta)p_L(\theta)},
\qquad
p_H(\theta)=(1-\rho)+(2\rho-1)\theta,
\qquad
p_L(\theta)=1-p_H(\theta).
\]
Differentiating and using $\sigma'(z)=\sigma(z)(1-\sigma(z))$ yields, for $\rho\in(1/2,1)$,
\[
|h'(z)|
=\bigl|\Delta'(\sigma(z))\bigr|\,\sigma'(z)
= r\,\frac{t(z)\,|2\sigma(z)-1|}{(1+t(z))^2}
\le \frac{r}{4}
\quad\text{for all }z\in\mathbb R,
\]
where $t(z)\triangleq \frac{(2\rho-1)^2\sigma(z)(1-\sigma(z))}{\rho(1-\rho)}$ and we used $x/(1+x)^2\le 1/4$ for all $x\ge 0$.
When $\rho=1$, we have $h(z)\equiv 2r-2$, so $|h'(z)|=0$ and the same bound holds. Therefore, $h$ is globally Lipschitz with constant $L:=r/4<1/2$.
For each $n$, define the update map $T_n(z):=z+\gamma_n h(z)$. For any $z\le z'$,
\[
T_n(z')-T_n(z)
=(z'-z)+\gamma_n\bigl(h(z')-h(z)\bigr)
\ge (1-\gamma_n L)(z'-z)
\ge 0,
\]
so $T_n$ is increasing.
In particular, if $h(\bar z)=0$ and $z_n\ge \bar z$, then $z_{n+1}=T_n(z_n)\ge T_n(\bar z)=\bar z$.

\medskip
\noindent\emph{Case 1: Low-fidelity regime ($\rho < \rho_c(r)$).}

By Proposition~\ref{prop:H-dominance}, $s(\rho, r) > 1$ implies $\Delta(\theta) < 0$ for all $\theta \in (0, 1)$. The limiting recursion satisfies
\[
z_{n+1} = z_n + \gamma_n \Delta(\theta_n) < z_n,
\]
so $(z_n)_{n \geq 0}$ is strictly decreasing. Since $\Delta(\theta)$ is bounded away from zero on any compact subset of $(0, 1)$ and $\sum_n \gamma_n = \infty$, we have $z_n \to -\infty$ and hence $\theta_n = \sigma(z_n) \to 0$.

\medskip
\noindent\emph{Case 2: Critical regime ($\rho = \rho_c(r)$).}

Here $s(\rho, r) = 1$, so $\Delta(1/2) = 0$ and $\Delta(\theta) < 0$ for all $\theta \in (0, 1) \setminus \{1/2\}$.

If $\theta_0 = 1/2$, then $\Delta(\theta_0) = 0$ implies $z_1 = z_0 + \gamma_0 \cdot 0 = z_0$, and by induction $z_n = z_0$ for all $n$, so $\theta_n \equiv 1/2$.

If $\theta_0 < 1/2$ (equivalently, $z_0 < 0$), then $\Delta(\theta_0) < 0$, so $z_1 < z_0 < 0$. Since $\Delta(\theta) < 0$ for all $\theta \neq 1/2$, the sequence $(z_n)$ remains negative and strictly decreasing. By the same argument as Case 1, $z_n \to -\infty$ and $\theta_n \to 0$.

If $\theta_0 > 1/2$ (equivalently, $z_0 > 0$), then $\Delta(\theta_0) < 0$, so $z_1<z_0$. Since $h(0)=\Delta(1/2)=0$ and each $T_n$ is increasing, $z_n\ge 0$ implies $z_{n+1}=T_n(z_n)\ge T_n(0)=0$. Thus $z_n\ge 0$ for all $n$.
Moreover, $h(z)<0$ for all $z>0$, so $(z_n)$ is strictly decreasing and bounded below by $0$, hence $z_n\to z^\star$ for some $z^\star\ge 0$. If $z^\star>0$, then $h(z^\star)<0$ and by continuity there exists $c>0$ such that $h(z)\le -c$ for all $z$ in a neighborhood of $z^\star$. For all $n$ large enough we then have $z_{n+1}\le z_n-c\gamma_n$, which implies $z_n\to -\infty$ since $\sum_n\gamma_n=\infty$, a contradiction. Therefore $z^\star=0$ and $\theta_n=\sigma(z_n)\to 1/2$.

\medskip
\noindent\emph{Case 3: High-fidelity regime ($\rho_c(r) < \rho < 1$).}

By Proposition~\ref{prop:H-dominance}, $s(\rho, r) < 1$ implies the existence of thresholds $\theta_-(\rho, r) < 1/2 < \theta_+(\rho, r)$ such that
\[
\Delta(\theta)=\begin{cases}
< 0, & \theta \in (0, \theta_-), \\
= 0, & \theta \in \{\theta_-, \theta_+\}, \\
> 0, & \theta \in (\theta_-, \theta_+), \\
< 0, & \theta \in (\theta_+, 1).
\end{cases}
\]

\emph{Subcase 3(a): $\theta_0 \in (0, \theta_-)$.} Then $z_0<z_-:=\log\!\bigl(\theta_-/(1-\theta_-)\bigr)$ and $h(z)<0$ for all $z<z_-$. Hence $(z_n)$ is strictly decreasing and $z_n\to -\infty$, so $\theta_n\to 0$.

\emph{Subcase 3(b): $\theta_0 \in (\theta_-, 1)$.} Then $z_0>z_-$, and since $h(z_-)=0$ and $T_n$ is increasing, we have $z_n>z_-$ for all $n$. Let $z_+:=\log\!\bigl(\theta_+/(1-\theta_+)\bigr)$. The sign pattern of $\Delta$ implies $h(z)>0$ for $z\in(z_-,z_+)$ and $h(z)<0$ for $z>z_+$. Therefore, if $z_n\le z_+$ then $z_{n+1}\ge z_n$, and if $z_n\ge z_+$ then $z_{n+1}\le z_n$. In either case, the recursion moves toward $z_+$.

If $z_n$ crosses $z_+$ only finitely many times, then eventually it remains on one side of $z_+$ and is monotone; hence, it converges to some limit $z^\star$. If $h(z^\star)\neq 0$, then by continuity there exists $c>0$ such that $|h(z_n)|\ge c$ for all $n$ large enough, which implies $|z_{n+1}-z_n|=\gamma_n|h(z_n)|\ge c\gamma_n$ eventually. Since $\sum_n\gamma_n=\infty$, this contradicts convergence of $(z_n)$. Therefore $h(z^\star)=0$, and given $z_n>z_-$ for all $n$ this forces $z^\star=z_+$.

If $z_n$ crosses $z_+$ infinitely many times, let $H:=\sup_{z\in\mathbb R}|h(z)|<\infty$. At each crossing between $n$ and $n+1$, the point $z_+$ lies between $z_n$ and $z_{n+1}$, so
\[
\max\{|z_n-z_+|,\ |z_{n+1}-z_+|\}\le |z_{n+1}-z_n|=\gamma_n|h(z_n)|\le \gamma_n H.
\]
Since $\gamma_n\to 0$, the distance to $z_+$ at crossing times tends to $0$. Between crossings, the recursion is monotone and stays between $z_+$ and the most recent crossing value, so the distance to $z_+$ is bounded by the corresponding crossing distance. Therefore $|z_n-z_+|\to 0$ and $z_n\to z_+$.

In both cases, $z_n\to z_+$ and therefore $\theta_n=\sigma(z_n)\to \theta_+(\rho,r)$.

\emph{Subcase 3(c): $\theta_0 = \theta_-$.} Then $\Delta(\theta_0) = 0$ implies $z_1 = z_0 + \gamma_0 \cdot 0 = z_0$, and by induction $z_n = z_0$ for all $n$, so $\theta_n \equiv \theta_-$.
This equilibrium is unstable. Given any neighborhood $U$ of $\theta_-$, choose $\varepsilon>0$ such that $\theta_- - \varepsilon \in U$ and $\theta_- + \varepsilon \in U$. Subcases 3(a) and 3(b) imply that the trajectory started from $\theta_- - \varepsilon$ converges to $0$, while the trajectory started from $\theta_- + \varepsilon$ converges to $\theta_+(\rho,r)$. Hence, arbitrarily small perturbations of $\theta_-$ move away from $\theta_-$.

\medskip
\noindent\emph{Case 4: Perfect-fidelity regime ($\rho = 1$).}

When $\rho = 1$, only the outcomes $(H, H)$ and $(L, L)$ occur (since recommendations match the latent mode with probability one). The conditional payoff difference becomes
\[
\Delta(\theta) = \mathbb{E}[\Pi \mid S = H] - \mathbb{E}[\Pi \mid S = L] = 2r - 2 > 0
\]
for all $\theta \in (0, 1)$, since $r > 1$.

Thus $z_{n+1} = z_n + \gamma_n(2r - 2) > z_n$ for all $n$, and since $\sum_n \gamma_n = \infty$, we have $z_n \to +\infty$ and $\theta_n \to 1$.

\medskip
This completes the proof of Proposition~\ref{prop:convergence-binf}.
\qed

\vskip 0.5cm
\noindent \textbf{Proof of Proposition \ref{prop:convergence-general}.}
Recall the stochastic approximation form \eqref{eq:SA-form-z}:
\[
z_{n+1} = z_n + \gamma_n\bigl(\Delta(\theta_n)+M_{n+1}\bigr),
\qquad \theta_n=\sigma(z_n).
\]
The step sizes $\gamma_n=\eta/(n+1)^\alpha$ satisfy $\sum_n\gamma_n=\infty$ and $\sum_n\gamma_n^2<\infty$ because $\alpha\in(1/2,1]$. For fixed $\rho\in(1/2,1)$, Lemma~\ref{lem:unbiased-D} implies $\mathbb{E}[M_{n+1}\mid \mathcal H_n]=0$ and $\sup_n \mathbb{E}[M_{n+1}^2]<\infty$.

\paragraph{Step 1: Noise summability.}
Define the martingale partial sums $W_n \triangleq \sum_{k=0}^{n-1}\gamma_k M_{k+1}$. Since $\sum_k\gamma_k^2<\infty$ and the conditional second moments of $M_{k+1}$ are uniformly bounded, $(W_n)$ is $L^2$ bounded and hence converges almost surely to a finite limit $W_\infty$ \citep[Theorem~5.4.9]{durrett2019}.
Consequently,
\begin{equation}\label{eq:z-decomposition-convergence-general}
z_n = z_0 + \sum_{k=0}^{n-1}\gamma_k\,\Delta(\theta_k) + W_n.
\end{equation}

\paragraph{Step 2: Excluding $\theta_n\to 1$.}
Because $\rho<1$, Proposition~\ref{prop:H-dominance} implies $\Delta(1)<0$. By continuity of $\Delta$ on $[0,1]$, there exist $\bar\theta\in(0,1)$ and $c_1>0$ such that $\Delta(\theta)\le -c_1$ for all $\theta\in[\bar\theta,1]$.
If $\theta_n\to 1$, then $\theta_n\ge \bar\theta$ for all large $n$, and \eqref{eq:z-decomposition-convergence-general} would imply
\[
z_n \le z_0 - c_1\sum_{k=0}^{n-1}\gamma_k + W_n \to -\infty,
\]
which contradicts $\theta_n=\sigma(z_n)\to 1$. Therefore, $\mathbb{P}(\theta_n\to 1)=0$.

\paragraph{Step 3: Convergence to ODE equilibria.}
Using the first-order expansion \eqref{eq:theta-firstorder} and the definition of $M_{n+1}$, we can write
\begin{equation}\label{eq:theta-SA-convergence-general}
\theta_{n+1} = \theta_n + \gamma_n\Bigl(\theta_n(1-\theta_n)\Delta(\theta_n) + \theta_n(1-\theta_n)M_{n+1}\Bigr) + r_{n+1},
\end{equation}
where the remainder satisfies $|r_{n+1}|\le C\gamma_n^2$ for a constant $C$ that depends only on $(\rho,r)$ and the payoff bounds. Since $\sum_n\gamma_n^2<\infty$, we also have $\sum_n |r_{n+1}|<\infty$.
The drift $g(\theta)\triangleq \theta(1-\theta)\Delta(\theta)$ is continuously differentiable on $[0,1]$ for $\rho\in(1/2,1)$, hence Lipschitz. Since $\theta_n\in(0,1)$ is bounded, standard ODE-method results for stochastic approximation applied to \eqref{eq:theta-SA-convergence-general} imply that $\theta_n$ converges almost surely to an equilibrium of the ODE \eqref{eq:ODE} \citep{borkar2008,benaim1999}.

\paragraph{Step 4: Identifying the possible limits.}
The equilibria of \eqref{eq:ODE} are $\theta\in\{0,1\}$ and any interior $\theta^\star$ satisfying $\Delta(\theta^\star)=0$. Proposition~\ref{prop:H-dominance} characterizes the interior zeros.
\begin{enumerate}
\item If $\rho\in(1/2,\rho_c(r))$, then $\Delta(\theta)<0$ for all $\theta\in(0,1)$, so the only equilibria are $\theta\in\{0,1\}$. Since $\mathbb{P}(\theta_n\to 1)=0$ by Step~2, we obtain $\theta_n\to 0$ almost surely.
\item If $\rho=\rho_c(r)$, then $\Delta(\theta)=0$ if and only if $\theta=1/2$, so the equilibria are $\theta\in\{0,1/2,1\}$. Step~2 rules out $\theta_n\to 1$, hence $\theta_\infty\in\{0,1/2\}$ almost surely.
\item If $\rho\in(\rho_c(r),1)$, then $\Delta(\theta)=0$ if and only if $\theta\in\{\theta_-(\rho,r),\theta_+(\rho,r)\}$, so the equilibria are $\theta\in\{0,\theta_-,\theta_+,1\}$. Step~2 rules out $\theta_n\to 1$. Moreover, $\theta_-(\rho,r)$ is linearly unstable for \eqref{eq:ODE}, and the noise in \eqref{eq:theta-SA-convergence-general} is nondegenerate on any neighborhood of $\theta_-$ for fixed $b$. Standard nonconvergence results for stochastic approximation then imply $\mathbb{P}(\theta_n\to \theta_-)=0$, so $\theta_\infty\in\{0,\theta_+(\rho,r)\}$ almost surely.
\end{enumerate}
\qed

\vskip 0.5cm
\noindent \textbf{Proof of Proposition \ref{prop:batch-selection}.}
Throughout, write $h(z)\triangleq \Delta(\sigma(z))$ so that the stochastic recursion in log-odds form is
\[
z_{n+1}=z_n+\gamma_n\bigl(h(z_n)+M_{n+1}\bigr),\qquad \theta_n=\sigma(z_n),
\]
where $M_{n+1}\triangleq \overline D_n^{(b)}-\Delta(\theta_n)$ is a martingale-difference term with $\mathbb{E}[M_{n+1}\mid \mathcal H_n]=0$.

\paragraph{Part (i).}
Fix $T>0$ and let $N(T)\triangleq \min\{n:\sum_{k=0}^{n-1}\gamma_k\ge T\}$.
Let $z_n^{\det}$ denote the deterministic recursion \eqref{eq:deterministic-update-z} and write $\theta_n^{\det}\triangleq \sigma(z_n^{\det})$.
Define the error $e_n\triangleq z_n-z_n^{\det}$, which satisfies
\[
e_{n+1}
=e_n+\gamma_n\bigl(h(z_n)-h(z_n^{\det})\bigr)+\gamma_n M_{n+1}.
\]
Define the martingale partial sums $W_m\triangleq \sum_{k=0}^{m-1}\gamma_k M_{k+1}$, with $W_0=0$, and set $u_n\triangleq e_n-W_n$.
Then
\[
u_{n+1}=u_n+\gamma_n\bigl(h(z_n)-h(z_n^{\det})\bigr).
\]
As shown in the proof of Proposition~\ref{prop:convergence-binf}, $h$ is globally Lipschitz on $\mathbb{R}$ with some constant $L<\infty$.
Let $W_\star\triangleq \max_{0\le m\le N(T)}|W_m|$.
For $n\le N(T)$,
\[
|u_{n+1}|
\le |u_n|+\gamma_n L |e_n|
=|u_n|+\gamma_n L |u_n+W_n|
\le (1+L\gamma_n)|u_n|+L\gamma_n W_\star.
\]
Iterating this inequality and using $\sum_{k=0}^{N(T)-1}\gamma_k\le T+\gamma_0\le T+1$ yields
\[
\max_{0\le n\le N(T)}|u_n|\le e^{L(T+1)}\,L(T+1)\,W_\star.
\]
Since $e_n=u_n+W_n$, we obtain
\[
\max_{0\le n\le N(T)}|e_n|
\le \bigl(1+e^{L(T+1)}L(T+1)\bigr)\,W_\star.
\]
Because $\overline D_n^{(b)}$ is a batch average of bounded per-round statistics (Lemma~\ref{lem:unbiased-D}), there exists $V<\infty$ depending only on $(\rho,r)$ such that
\[
\mathbb{E}\bigl[M_{n+1}^2\mid \mathcal H_n\bigr]\le \frac{V}{b}\qquad\text{a.s. for all }n.
\]
Therefore,
\[
\mathbb{E}[W_{N(T)}^2]
=\sum_{k=0}^{N(T)-1}\gamma_k^2\,\mathbb{E}[M_{k+1}^2]
\le \frac{V}{b}\sum_{k=0}^{N(T)-1}\gamma_k^2.
\]
By Doob's $L^2$ maximal inequality,
\[
\mathbb{P}\bigl(W_\star>\xi\bigr)
\le \frac{4\,\mathbb{E}[W_{N(T)}^2]}{\xi^2}
\le \frac{4V\sum_{k=0}^{N(T)-1}\gamma_k^2}{b\,\xi^2}.
\]
Using that $\sigma$ is $1/4$ Lipschitz, we have $|\theta_n-\theta_n^{\det}|\le \frac{1}{4}|e_n|$, hence
\[
\mathbb{P}\Bigl(\max_{0\le n\le N(T)}|\theta_n-\theta_n^{\det}|>\varepsilon\Bigr)
\le \frac{C(T,\rho,r,\alpha,\eta)}{b\,\varepsilon^2}.
\]
Passing from discrete time to the interpolations $\bar\theta(t)$ and $\theta^{\det}(t)$ on $t\in[0,T]$ changes only constants, because the interpolation error is $O(\max_{n\le N(T)}\gamma_n)$ and does not depend on $b$.
This yields the displayed bound in Part (i) and the $O_p(1/\sqrt b)$ statement follows.

\paragraph{Part (ii).}
Fix $\delta\in(0,1/2)$ and recall
\[
K_+(\delta)\triangleq[\theta_-+\delta,\,1-\delta],
\qquad
K_-(\delta)\triangleq[\delta,\,\theta_- - \delta].
\]
Write
\[
z_-:=\log\!\Bigl(\frac{\theta_-}{1-\theta_-}\Bigr),
\qquad
z_+:=\log\!\Bigl(\frac{\theta_+}{1-\theta_+}\Bigr).
\]
By Proposition~\ref{prop:convergence-general}, in this regime $\theta_n$ converges almost surely, and the only possible limits are $0$ and $\theta_+(\rho,r)$.

\smallskip
\noindent
\emph{Step 1: A maximal inequality for the weighted martingale noise.}
Write the batch estimator as an average of $b$ i.i.d.\ per-round scores.
Define
\[
\widetilde D_n^{(\ell)}\triangleq \frac{1}{2}\sum_{i=1}^2 D_i^{(\ell)}(\theta_n),
\qquad \ell=1,\dots,b,
\]
so that $\overline D_n^{(b)}=\frac{1}{b}\sum_{\ell=1}^b \widetilde D_n^{(\ell)}$.
Conditional on $\mathcal H_n$, the random variables $(\widetilde D_n^{(\ell)})_{\ell=1}^b$ are i.i.d.\ with mean $\Delta(\theta_n)$.
By Lemma~\ref{lem:unbiased-D}, there exist constants $\underline D<\overline D$ depending only on $(\rho,r)$ such that
$\underline D\le \widetilde D_n^{(\ell)}\le \overline D$ almost surely for all $n$ and $\ell$.
Consequently, Hoeffding's lemma implies that for all $\lambda\in\mathbb R$,
\begin{equation}\label{eq:batch-selection-subg}
\mathbb E\!\left[\exp(\lambda M_{n+1})\mid \mathcal H_n\right]
\le
\exp\!\left(\frac{\lambda^2 V_\star}{2b}\right),
\qquad
V_\star:=\Bigl(\frac{\overline D-\underline D}{2}\Bigr)^2.
\end{equation}

Define the weighted martingale
\[
W_n:=\sum_{k=0}^{n-1}\gamma_k M_{k+1},
\qquad
W_0=0,
\qquad
S_2:=\sum_{k=0}^\infty\gamma_k^2<\infty.
\]
Fix $\lambda>0$ and define
\[
\mathcal E_n(\lambda):=\exp\!\left(\lambda W_n-\frac{\lambda^2 V_\star}{2b}\sum_{k=0}^{n-1}\gamma_k^2\right).
\]
Using \eqref{eq:batch-selection-subg} and iterated expectations, $(\mathcal E_n(\lambda),\mathcal H_n)$ is a nonnegative supermartingale with $\mathbb E[\mathcal E_0(\lambda)]=1$.
By Ville's inequality,
\[
\mathbb P\!\left(\sup_{n\ge 0} W_n\ge t\right)
\le
\exp\!\left(-\lambda t+\frac{\lambda^2 V_\star S_2}{2b}\right).
\]
Optimizing over $\lambda$ yields
\[
\mathbb P\!\left(\sup_{n\ge 0} W_n\ge t\right)\le \exp\!\left(-\frac{b\,t^2}{2V_\star S_2}\right),
\qquad
\mathbb P\!\left(\inf_{n\ge 0} W_n\le -t\right)\le \exp\!\left(-\frac{b\,t^2}{2V_\star S_2}\right).
\]
Therefore, for every $t>0$,
\begin{equation}\label{eq:batch-selection-W-max}
\mathbb P\!\left(\sup_{n\ge 0}|W_n|\ge t\right)
\le
2\exp\!\left(-\frac{b\,t^2}{2V_\star S_2}\right).
\end{equation}

\smallskip
\noindent
\emph{Step 2: A deterministic barrier lemma.}
Define the down-crossing and up-crossing times of the unstable log-odds threshold $z_-$,
\[
\tau_-:=\inf\{n\ge 0:z_n\le z_-\},
\qquad
\tau_+:=\inf\{n\ge 0:z_n\ge z_-\}.
\]

\noindent\emph{Claim (Barrier bound for separatrix crossings).}
Fix $\delta\in(0,1/2)$ such that $\theta_- - \delta$ and $\theta_-+\delta$ lie in $(0,1)$, and define
\[
z_\delta^+:=\log\!\Bigl(\frac{\theta_-+\delta}{1-(\theta_-+\delta)}\Bigr),
\qquad
z_\delta^-:=\log\!\Bigl(\frac{\theta_- - \delta}{1-(\theta_- - \delta)}\Bigr),
\qquad
z_{\mathrm{mid}}:=\frac{z_-+z_+}{2},
\qquad
\underline z_\delta:=z_\delta^+\wedge z_{\mathrm{mid}}.
\]
\begin{enumerate}[(a)]
\item If $z_0\ge z_\delta^+$ and $\tau_-<\infty$, then $\sup_{n\ge 0}|W_n|\ge (\underline z_\delta-z_-)/2$.
\item If $z_0\le z_\delta^-$ and $\tau_+<\infty$, then $\sup_{n\ge 0}|W_n|\ge (z_- - z_\delta^-)/2$.
\end{enumerate}

\emph{Proof.}
\emph{Part (a).}
Assume $z_0\ge z_\delta^+$ and $\tau_-<\infty$.
Define the last time before $\tau_-$ at which the process is at or above $\underline z_\delta$,
\[
\tau_\delta:=\sup\{n\le \tau_-: z_n\ge \underline z_\delta\}.
\]
Then $z_{\tau_\delta}\ge \underline z_\delta$ and $z_{\tau_\delta+1}<\underline z_\delta$.

If $z_{\tau_\delta}\le z_+$, then for every $k$ with $\tau_\delta\le k<\tau_-$ we have $z_-<z_k\le z_+$ and hence $h(z_k)\ge 0$.
Writing $z_n=z_0+\sum_{k=0}^{n-1}\gamma_k h(z_k)+W_n$ gives
\[
z_{\tau_-}
=
z_{\tau_\delta}
+\sum_{k=\tau_\delta}^{\tau_- -1}\gamma_k h(z_k)
+\bigl(W_{\tau_-}-W_{\tau_\delta}\bigr).
\]
On $\{\tau_-<\infty\}$, the middle term is nonnegative, so
\[
W_{\tau_-}-W_{\tau_\delta}
\le
z_{\tau_-}-z_{\tau_\delta}
\le
z_- - \underline z_\delta
=-(\underline z_\delta-z_-).
\]
If $\sup_n|W_n|<(\underline z_\delta-z_-)/2$, then $W_{\tau_\delta}\le \sup_n|W_n|<(\underline z_\delta-z_-)/2$ and hence
\[
W_{\tau_-}
\le
W_{\tau_\delta}-(\underline z_\delta-z_-)
<
\frac{\underline z_\delta-z_-}{2}-(\underline z_\delta-z_-)
=
-\frac{\underline z_\delta-z_-}{2},
\]
which contradicts $\sup_n|W_n|<(\underline z_\delta-z_-)/2$.

If instead $z_{\tau_\delta}> z_+$, define $T_n(z):=z+\gamma_n h(z)$.
As shown in the proof of Proposition~\ref{prop:convergence-binf}, $|h'(z)|\le r/4$ for all $z$ and since $\gamma_n\le 1$ this implies $T_n$ is increasing.
Because $T_{\tau_\delta}(z_+)=z_+$ and $z_{\tau_\delta}\ge z_+$, we obtain $T_{\tau_\delta}(z_{\tau_\delta})\ge z_+$.
Therefore,
\[
\bigl|W_{\tau_\delta+1}-W_{\tau_\delta}\bigr|
=
\bigl|z_{\tau_\delta+1}-T_{\tau_\delta}(z_{\tau_\delta})\bigr|
\ge
z_+-\underline z_\delta.
\]
Since $\underline z_\delta\le z_{\mathrm{mid}}=(z_-+z_+)/2$, we have $z_+-\underline z_\delta\ge \underline z_\delta-z_-$.
Using
$|W_{\tau_\delta+1}-W_{\tau_\delta}|\le |W_{\tau_\delta+1}|+|W_{\tau_\delta}|\le 2\sup_n|W_n|$
yields $\sup_n|W_n|\ge (\underline z_\delta-z_-)/2$.

\smallskip
\emph{Part (b).}
Assume $z_0\le z_\delta^-$ and $\tau_+<\infty$.
Define the last time before $\tau_+$ at which the process is at or below $z_\delta^-$,
\[
\widehat\tau_\delta:=\sup\{n\le \tau_+: z_n\le z_\delta^-\}.
\]
Then $z_{\widehat\tau_\delta}\le z_\delta^-$, and for every $k$ with $\widehat\tau_\delta\le k<\tau_+$ we have $z_\delta^-<z_k<z_-$, hence $h(z_k)\le 0$.
Decomposing again,
\[
z_{\tau_+}
=
z_{\widehat\tau_\delta}
+\sum_{k=\widehat\tau_\delta}^{\tau_+ -1}\gamma_k h(z_k)
+\bigl(W_{\tau_+}-W_{\widehat\tau_\delta}\bigr),
\]
and the middle term is nonpositive, so
\[
W_{\tau_+}-W_{\widehat\tau_\delta}
\ge
z_{\tau_+}-z_{\widehat\tau_\delta}
\ge
z_- - z_\delta^-.
\]
If $\sup_n|W_n|<(z_- - z_\delta^-)/2$, then
$W_{\widehat\tau_\delta}\ge -\sup_n|W_n|>-(z_- - z_\delta^-)/2$
and thus
\[
W_{\tau_+}
\ge
W_{\widehat\tau_\delta}+(z_- - z_\delta^-)
>
-\frac{z_- - z_\delta^-}{2}+(z_- - z_\delta^-)
=
\frac{z_- - z_\delta^-}{2},
\]
contradicting $\sup_n|W_n|<(z_- - z_\delta^-)/2$.
\hfill$\square$

\smallskip
\noindent
\emph{Step 3: Conclude the exponential bounds.}
Fix $\delta\in(0,1/2)$.

If $\theta_0\in K_+(\delta)$ then $z_0\ge z_\delta^+$.
On the event $\{\theta_n\to 0\}$ we have $z_n\to-\infty$, hence $\tau_-<\infty$.
Therefore, using Step~2(a) and \eqref{eq:batch-selection-W-max},
\[
1-p_+(b,\theta_0)
=\mathbb P(\theta_n\to 0\mid \theta_0)
\le
\mathbb P\!\left(\sup_{n\ge 0}|W_n|\ge \frac{\underline z_\delta-z_-}{2}\ \middle|\ \theta_0\right)
\le
2\exp\!\left(-\frac{b\,(\underline z_\delta-z_-)^2}{8V_\star S_2}\right).
\]
Similarly, if $\theta_0\in K_-(\delta)$ then $z_0\le z_\delta^-$.
On the event $\{\theta_n\to\theta_+\}$ we have $z_n\to z_+>z_-$, hence $\tau_+<\infty$.
Step~2(b) and \eqref{eq:batch-selection-W-max} give
\[
p_+(b,\theta_0)
\le
\mathbb P\!\left(\sup_{n\ge 0}|W_n|\ge \frac{z_- - z_\delta^-}{2}\ \middle|\ \theta_0\right)
\le
2\exp\!\left(-\frac{b\,(z_- - z_\delta^-)^2}{8V_\star S_2}\right).
\]
Finally, note that
\[
z_\delta^+-z_-=\int_{\theta_-}^{\theta_-+\delta}\frac{\diff\theta}{\theta(1-\theta)}\ge 4\delta,
\qquad
z_- - z_\delta^-=\int_{\theta_- - \delta}^{\theta_-}\frac{\diff\theta}{\theta(1-\theta)}\ge 4\delta,
\]
and $\underline z_\delta-z_-=(z_\delta^+-z_-)\wedge((z_+-z_-)/2)$.
This implies the displayed bounds in Part (ii) after absorbing constants into $c_\delta$ and $C_\delta$.
For any fixed $\theta_0\ne\theta_-$, choose $\delta>0$ such that $\theta_0\in K_+(\delta)$ when $\theta_0>\theta_-$ and $\theta_0\in K_-(\delta)$ when $\theta_0<\theta_-$.
Letting $b\to\infty$ then yields $\lim_{b\to\infty}p_+(b,\theta_0)=\mathbbm{1}\{\theta_0>\theta_-\}$.

	\paragraph{Part (iii).}
	Fix $\epsilon\in(0,1/2)$ and $\delta_0\in(0,1/2)$ with $\theta_-\in(\delta_0,1-\delta_0)$.
	Set
	\[
	\kappa_{\delta_0}:=\min\!\left\{4,\frac{z_+-z_-}{2\delta_0}\right\}.
	\]
	Let $\theta_0\in[\delta_0,1-\delta_0]$ and define $\delta:=\min\{\tfrac12|\theta_0-\theta_-|,\delta_0\}$.
	If $\theta_0=\theta_-$, the conclusion is immediate, so assume $\theta_0\ne\theta_-$.
	If $\theta_0>\theta_-$ then $\theta_0\in K_+(\delta)$, and if $\theta_0<\theta_-$ then $\theta_0\in K_-(\delta)$.

For $\delta\le \delta_0$ we have
\[
\underline z_\delta-z_-=(z_\delta^+-z_-)\wedge\Bigl(\frac{z_+-z_-}{2}\Bigr)\ge \kappa_{\delta_0}\,\delta,
\qquad
z_- - z_\delta^-\ge 4\delta\ge \kappa_{\delta_0}\,\delta.
\]
Therefore, by the bounds established in Part (ii), for all $b\ge 1$,
	\[
	p_+(b,\theta_0)\in(\epsilon,1-\epsilon)
	\Longrightarrow
	\epsilon\le 2\exp\!\left(-\frac{\kappa_{\delta_0}^2}{8V_\star S_2}\,b\,\delta^2\right).
	\]
	Rearranging yields
	\[
	\delta^2\le \frac{8V_\star S_2}{\kappa_{\delta_0}^2 b}\log\!\Bigl(\frac{2}{\epsilon}\Bigr).
	\]
	Define
	\[
	A:=\sqrt{\frac{8V_\star S_2}{\kappa_{\delta_0}^2}\log\!\Bigl(\frac{2}{\epsilon}\Bigr)}.
	\]
	Then $\delta\le A/\sqrt b$.
	If $|\theta_0-\theta_-|\le 2\delta_0$, we have $\delta=\tfrac12|\theta_0-\theta_-|$, so $|\theta_0-\theta_-|=2\delta\le 2A/\sqrt b$.
	If instead $|\theta_0-\theta_-|>2\delta_0$, we have $\delta=\delta_0$, so $\delta_0\le A/\sqrt b$ and hence $b\le A^2/\delta_0^2$.
	Since $|\theta_0-\theta_-|\le 1$, this implies $|\theta_0-\theta_-|\le (A/\delta_0)/\sqrt b$.
	Taking $K:=A/\delta_0$ completes the proof.
	\qed

\end{document}